\documentclass[12pt]{article}
\usepackage{amsfonts}
\usepackage{mathrsfs}
\usepackage{amsthm}
\usepackage{multirow}
\usepackage{bbding}
\usepackage{amssymb}
\usepackage{amsmath}
\usepackage{graphicx,color}
\usepackage{rotating}
\usepackage{bm}
\usepackage{color}
\newcommand{\bea}{\begin{eqnarray}}
\newcommand{\eea}{\end{eqnarray}}
\newcommand{\Bea}{\begin{eqnarray*}}
\newcommand{\Eea}{\end{eqnarray*}}
\newcommand{\ba}{\begin{array}}
\newcommand{\ea}{\end{array}}
\newcommand{\bt}{\begin{tabular}}
\newcommand{\et}{\end{tabular}}
\newcommand{\btb}{\begin{table}}
\newcommand{\etb}{\end{table}}
\newcommand{\bc}{\begin{center}}
\newcommand{\ec}{\end{center}}
\newcommand{\beq}{\begin{equation}}
\newcommand{\eeq}{\end{equation}}

\usepackage{float}

\makeatletter

\newcommand{\Rmnum}[1]{\expandafter\@slowromancap\romannumeral #1@}
\makeatother

\begin{document}

\title{A Global Bias-Correction DC Method for Biased Estimation under Memory Constraint }
\author{Lu Lin$^1$ and Feng Li$^2$\footnote{The corresponding
author. Email: lifengstat@zzu.edu.cn. The research was
supported by NNSF projects (11571204, U1404104) of China.}
\\
$^1$Zhongtai Securities Institute for Financial Studies\\ Shandong University, Jinan, China\\
$^2$School of Mathematics and Statistics\\
 Zhengzhou University, Zhengzhou, China
}
\date{}
\maketitle

\vspace{-0.5cm}

\begin{abstract} \baselineskip=15pt

This paper establishes a global bias-correction divide-and-conquer (GBC-DC) rule for biased estimation under the case of memory constraint. In order to introduce the new estimation, a closed representation of the local estimators obtained by the data in each batch is adopted, aiming to formulate a pro forma linear regression between the local estimators and the true parameter of interest. Least square method is then used within this framework to composite a global estimator of the parameter.
Thus, the main advantage over the classical DC method is that the new GBC-DC method can absorb the information hidden in the statistical structure and the variables in each batch of data. Consequently, the resulting global estimator is strictly unbiased even if the local estimator has a non-negligible bias. Moreover, the global estimator is consistent, and even can achieve root-$n$ consistency, without the constraint on the number of batches. Another attractive feature of the new method is computationally simple and efficient, without use of any iterative algorithm and local bias-correction.
Specifically, the proposed GBC-DC method applies to various biased estimations such as shrinkage-type estimation and nonparametric regression estimation. Detailed simulation studies demonstrate that the proposed GBC-DC approach is significantly bias-corrected, and the behavior is comparable with the full data estimation and is much better than the competitors.


\end{abstract}

\baselineskip=20pt

\setcounter{equation}{0}
\section{Introduction}
The divide-and-conquer (DC) in computer science is one of the most important algorithms
to deal with large-scale datasets. When
large-scale datasets cannot be fit into the memory of a single computer, they are distributed in many machines over
limited memory. Then, the local result (e.g., the local estimator of a parameter) can be obtained by
the batch of data in each machine, and finally, the global result can be achieved by aggregating these local results. See, e.g., Manku, Rajagopalan and Lindsay (1998); Greenwald and
Khanna (2004); Zhang and Wang (2007); Guha and Mcgregor (2009) and
the references therein.
Up to now, there have been various types of aggregation methodologies for constructing the global estimator, for instance, the naive average of the local estimators (see, e.g., Mcdonald et al. 2009; Zinkevich et al. 2010), and the relevant DC expressions (see, e.g., Chen, et al. 2006, and Lin and Xi, 2011) and representative approaches (see, e.g., Li and Yang, 2018, Wang, 2018). The related works include but are not
limited to the DC expression for linear model of Chen et al. (2006), Lin and Xi (2011), and Schifano et al. (2016), the density estimation of Li, Lin and Li (2013), the parametric regression estimation of Chen and Xie (2014), and Zhang, Duchi and Wainwright (2015), the high-dimensional parametric regression estimation of Lee et al. (2017), semi-parametric regression estimation of Zhao,
Cheng and Liu (2016), quantile regression processes of Volgushev, Chao and Cheng (2018), the $M$-estimator of Shi, Lu and Song (2017), and the distributed testing and estimation of Battey et al. (2018).

As suggested by the existing literature (see, e.g., Li,  Lin and Li, 2013; Zhang et al. 2013; Rosenblatt and Nadler 2016), for
achieving the same asymptotic distribution for statistical inference as pooling all the data
together, the number of batches is restricted. More specifically, a commonly used restriction is $N=o(\sqrt n)$ (or equivalently $n=o(m^2)$), where $n$ is the sample size, $N$ is the number of batches and $m=n/N$. Such a constraint on $N$ cannot be satisfied in some applications such as sensor networks and streaming data, because the number of batches can be large.
In order to relax the constraint, instead of one-shot aggregation via averaging, the aggregation with multiple rounds (e.g., iterative algorithm) was proposed recently by Jordan, Lee and Yang (2018) and Wang et al. (2017) for the case of differentiable loss function, and Chen, Liu and Zhang (2018) for quantile regression with
non-differentiable loss. These methods are able to reduce both estimation bias and variance simultaneously and then obtain the standard result as pooling all the data
together. It is known that bias reduction is more crucial than the variance reduction. Such a goal cannot be achieved by many classical
inference methods that require to balance the variance and bias.

The estimation bias often appears in the procedure of statistical inference. The common examples are  shrinkage-type estimations in linear and generalized linear models, $M$- and $Z$-estimations in nonlinear regression model, and  kernel estimation in nonparametric regression model.
It is verified by our motivating examples in the next section that when the local estimator is biased (e.g., LASSO estimator), the global estimator by the naive average or the original DC expression cannot achieve $\sqrt n$-consistent and is even divergent when $N$ is large.
Thus, bias-correction has been considered in the existing DC literature. The most common procedures use iterative algorithm
(see, e.g., Wang et al., 2017) and local bias-correction (see, e.g., Lee, et al. 2017; Lian, et al., 2018; and Keren and Yang, 2018) to reduce the bias of local estimators and then to control the bias of the global estimator. However, the iterative algorithm and the bias-correction for local estimators are computationally complex, and the resulting bias-correction for global estimation is not sufficient.

From a new perspective, we in this paper explore a global bias-correction divide-and-conquer (GBC-DC, for short) algorithm for the biased local estimations under the case of large sample size.
The newly proposed GBC-DC methodology is motivated by a proven statistical technique, composition, which has received
much attention in the literature. The early goal of the classical composition methods is only to
reduce the estimation variance via optimizing the composite estimation covariance;
see Zou and Yuan (2008) for composite
quantile linear regression estimation, see Kai, Li, and Zou
(2010), and Sun, Gai, and Lin (2013) for composite nonparametric
regression estimation, see Kai, Li, and Zou (2011) for composite semiparametric estimation, see Bradic, Fan, and Wang (2011) for composite variable selection of ultra-high-dimensional
models. Recently, bias-reduction by composition has attained much attention as well in the literature. Based on the asymptotic or approximate representation of the initial estimator, Lin et al. (2019), Cheng et al. (2018) and Lin and Li (2008) introduced composite least squares to realize the targets of reducing estimation bias and optimizing estimation covariance, simultaneously. Moreover, the relevant composition methods were suggested by Wang et al. (2019), Dai et al. (2016 and 2017), Wang and Lin (2015), and Tong and Wang (2005) for constructing the composite estimators of the derivative and variance in nonparametric regression.

It will be seen later that the main advantage over the aggregation of DC in computer science is that the GBC-DC technique is able to sufficiently absorb the information of statistical structure and the variables in batches of data.
To realize our goals aforementioned, we employ a closed representation of the local estimator computed on each batch of data to build a pro forma linear regression model, in which the combination of the variables in each batch is regarded as the covariate and the local estimator is thought of as as response variable. Based on such a model and least squares, we composite a global estimator. It will be shown in the later development that this method has the following salient features.
\vspace{-1ex}
\begin{enumerate}
\item[1)] {\it Global bias-correction}. The new composition method sufficiently employs the information of the closed representation and the batches such that the resulting global estimator is strictly unbiased even if the local estimators have a non-negligible bias.
 \item[2)] {\it Acceleration of convergence}. The convergence rate of the global estimator is accelerated such that the $\sqrt n$-consistency can be achieved for any choices of $N$ and $m$.
\item[3)] {\it Simplicity}. Iterative algorithm and bias-correction for local estimators in the aggregation procedure are not needed. Furthermore, the structure of the resultant global estimator is simple, which is a least squares estimator and has a DC expression. Thus, the composition procedure is computationally simple and efficient. Benefiting from the structure of least squares, we can construct its online updating version and make statistical inference in the case of data streams.
    \item[4)] {\it Generality}. Although our method focuses mainly on linear model and related parameter estimations, the new technique is also extended into other models such as nonlinear and nonparametric models.
   \item[5)] {\it Innovation}. The use of the DC expressed model, instead of DC expressed estimation, is our main innovation.
\end{enumerate}
All the salient features above will be illustrated by our comprehensive
simulation studies, which particularly show that the global estimator by GBC-DC is significantly bias-corrected, and its behavior is much better than the competitors and is comparable with the full data estimation.

The remainder of this paper is organized in the following way. In Section 2, after the classical DC algorithm is briefly recalled, some motivating examples are investigated to motivate the methodological development. In Section 3, a unified framework for linear model is defined, and the bias-corrected global estimator is proposed via the newly defined model and least square method, and the theoretical properties of the global estimator are investigated. The extensions of the new method to the cases of nonlinear and nonparametric models are discussed in Section 4. Simulation studies are provided in Section 5 to illustrate the new method. The proofs of theorems are relegated to Appendix.

\setcounter{equation}{0}
\section{ Problem Formulation}

\subsection{Divide-and-conquer}

We briefly recall general DC algorithm for statistical
estimation. Let
$\{Z_1,\cdots, Z_n\}$ be the set of observation data, where the sample size $n$ is extremely large. Our goal is to estimate a $p$-dimensional parameter $\theta=(\theta^1,\cdots,\theta^p)^T$.
We split the data index set $\{1,\cdots, n\}$ into $N$ subsets ${\cal H}_1,\cdots,{\cal H}_N$, where the size of ${\cal H}_j$ is $m=|{\cal H}_j|$ satisfying $n= Nm$. Correspondingly, the entire
dataset $\{Z_1,\cdots, Z_n\}$ is divided into $N$ batches ${\cal D}_1,\cdots,{\cal D}_N$ with ${\cal D}_j=\{Z_i,i\in{\cal H}_j\}$. By swapping each batch of data
${\cal D}_j$ into the memory, we can construct a local estimator of $\theta$ as $\widehat\theta_j = g_j({\cal D}_j)$ for ${\cal D}_j$ with some function $g_j(\cdot)$. The global estimator $\widehat\theta$ is then obtained by an
aggregation of $\widehat\theta_j,
j=1,\cdots,N$, e.g., the naive average as $\widehat\theta=\frac 1N\sum_{j=1}^N\widehat\theta_j$ or the corresponding DC expression (see the motivating examples below). Actually, the classical DC strategy typically requires a
random data partition, that is, the batches of data stored in different computers are independent
and have the same distribution. In our setting, however, the identical distribution assumption on ${\cal D}_1,\cdots,{\cal D}_N$ is not necessary. We particularly consider the example of streaming data where the obtained data may not be identically distributed in different observation periods.

In this section, we mainly focuses on the following linear model: \begin{eqnarray}\label{(2.1)}Y_i=X_i^T\beta+\varepsilon_i,\ i=1,\cdots,n,\end{eqnarray} where $\beta=(\beta^1,\cdots,\beta^p)^T$ is a $p$-dimensional vector of unknown parameters, and $X_i=(X_i^1,\cdots,X_i^p)^T$, $i=1,\cdots,n$, are independent observations of a $p$-dimensional covariate $X=(X^1,\cdots,X^p)^T$, and the errors $\varepsilon_i,i=1,\cdots,n$, are independent and satisfy $E[\varepsilon_i|X_i]=0$ and $Var[\varepsilon_i|X_i]=\sigma_\varepsilon^2$. For the regression model, the data batches are ${\cal D}_j=\{(X_i,Y_i),i\in{\cal H}_j\},j=1,\cdots,N$.

\subsection{Motivating examples and related issues}

To proceed with the methodological development, we first review the following shrinkage-type estimators, their estimation biases and the related closed representations.

{\it Example 1. LASSO estimator.}
When the dimension $p$ is high in model (\ref{(2.1)}), we use penalty-based methods to select variables and estimate parameters, simultaneously. Based on the subset ${\cal D}_j$, the LASSO estimator (Tibshirani, 1996) of $\beta$ is given by
$$\widehat\beta_j=\arg\min_{\beta}\frac{1}{2m}\sum_{i\in{\cal H}_j}(Y_i-X_i^T\beta)^2+\lambda_j\|\beta\|_1,$$ where $\lambda_j>0$ is a regularization parameter satisfying
\begin{enumerate}
\item[{\it C0}.] $\lambda_j=O(m^{-\delta})$ for some constant $0<\delta\leq 1$. \end{enumerate}
For the condition, see, e.g., Knight and Fu (2000).
Without loss of generality, suppose that $\beta^{k}\neq 0$ for $k=1,\cdots,s$, and $\beta^{k}= 0$ for $k=s+1,\cdots,p$. Denote by $\beta_S$ the significant subset of $\beta$, i.e., $\beta_S=(\beta^1,\cdots,\beta^s)^T$. Let $X_S=(X^1,\cdots,X^s)^T$, ${\bf X}_S=({\bf x}_1,\cdots,{\bf x}_s)$ with ${\bf x}_k=(X_1^k,\cdots,X_n^k)^T$, and ${\bf X}_{jS}=({\bf x}_{j1},\cdots,{\bf x}_{js})$ with ${\bf x}_{jk}=(X_l^k:l\in{\cal H}_j)^T$.
The existing literature (e.g. Wainwright, 2009; Huang et al, 2008) reported that under some regularity conditions, the resultant estimator $\widehat \beta_{jS}$ of the significant subset $\beta_S$ has the following closed representation:
\begin{eqnarray}\label{(2.2)}\widehat \beta_{jS}=\beta_S-\left(\frac{1}{m}{\bf X}_{jS}^T{\bf X}_{jS}\right)^{-1}\lambda_j\mbox{sgn}(\beta_S)+\left(\frac{1}{m}{\bf X}_{jS}^T{\bf X}_{jS}\right)^{-1}\frac{1}{m}{\bf X}_{jS}^T{\bm \varepsilon}_j,\end{eqnarray} where ${\bm \varepsilon}_j=(\varepsilon_i:i\in{\cal H}_j)^T$. The above representation will be useful for our modeling, but now we mainly focus on the estimation bias. From (\ref{(2.2)}) we can see that the estimator is shrunken and has the estimation bias as $-E\left[\left(\frac{1}{m}{\bf X}_{jS}^T{\bf X}_{jS}\right)^{-1}\right]\lambda_j\mbox{sgn}(\beta_S)$.
Then, the naive average $\widehat\beta_{S}=\frac 1N\sum_{j=1}^N\widehat \beta_{jS}$ has the bias as $$B(\widehat\beta_{S})=-\frac1N\sum_{j=1}^NE\left[\left(\frac{1}{m}{\bf X}_{jS}^T{\bf X}_{jS}\right)^{-1}\right]\lambda_j\,\mbox{sgn}(\beta_S),$$ which is of order $O(m^{-\delta})$. Similarly, the DC expression of LASSO estimator
\begin{eqnarray}\label{(2.3)}\widehat \beta_{S}=\left(\sum_{j=1}^N\frac{1}{m}{\bf X}_{jS}^T{\bf X}_{jS}\right)^{-1}\sum_{j=1}^N\frac{1}{m}{\bf X}_{jS}^T{\bf X}_{jS}\widehat \beta_{jS}\end{eqnarray} (see, e.g., Lin and Xi, 2011) has the bias of order $O(m^{-\delta})$ as well. Thus, under Condition {\it C0}, $\sqrt n\,\widehat\beta_{S}$ has a bias of order $O(n^{1/2-\delta}N^{\delta}),$ and satisfies
\begin{eqnarray}\label{(2.4)}\sqrt nBias(\widehat\beta_{S})=O(n^{1/2-\delta}N^{\delta})\rightarrow\infty \mbox{ if } m^{2\delta-1}=o(N).\end{eqnarray}
This shows that the global estimator $\widehat\beta_{S}$ cannot achieve $\sqrt n$-consistency when $m^\delta=o(N^{1/2}m^{1/2})$. If $0<\delta\leq 1/2$, the condition $m^{2\delta-1}=o(N)$ always holds; when $1/2<\delta<1$, the condition $m^{2\delta-1}=o(N)$ means that $N$ should be larger than $m^{2\delta-1}$.


{\it Example 2. Ridge estimator.} Under model (\ref{(2.1)}), the Ridge estimator computed on subset ${\cal D}_j$ is defined by
$$\widehat\beta_j=\arg\min_{\beta}\frac{1}{m}\sum_{i\in{\cal H}_j}(Y_i-X_i^T\beta)^2+\lambda_j\|\beta\|_2^2.$$
Let ${\bf X}=({\bf x}_1,\cdots,{\bf x}_p)$ with ${\bf x}_k=(X_{1k},\cdots,X_{nk})^T$ and ${\bf X}_{j}=({\bf x}_{j1},\cdots,{\bf x}_{jp})$ with ${\bf x}_{jk}=(X^k_{l}:l\in{\cal H}_j)^T$.
It can be verified that the Ridge estimator has the following closed representation:
\begin{eqnarray}\label{(2.5)}\widehat \beta_{j}=\beta-\left(\frac{1}{m}{\bf X}_{j}^T{\bf X}_{j}+\lambda_j I_p\right)^{-1}\lambda_j\beta +\left(\frac{1}{m}{\bf X}_{j}^T{\bf X}_{j}+\lambda_j I_p\right)^{-1}\frac{1}{m}{\bf X}_{j}^T{\bm \varepsilon}_j,\end{eqnarray}
where $I_p$ is a $p\times p$ identity matrix. Similar to (\ref{(2.2)}), the above representation will be useful for our modeling, but now we mainly focus on the estimation bias as well.
The estimator is shrunken and has the estimation bias of order $O(m^{-\delta})$.
Then, the naive average $\widehat\beta=\frac 1N\sum_{j=1}^N\widehat \beta_{j}$ has the estimation bias as $$B(\widehat\beta)=-\frac1N\sum_{j=1}^NE\left[\left(\frac{1}{m}{\bf X}_{j}^T{\bf X}_{j}+\lambda_j I_p\right)^{-1}\right]\lambda_j\beta,$$ which is of order $O(m^{-\delta})$.  Similarly, the DC expression of Ridge  estimator
\begin{eqnarray}\label{(2.6)}\widehat \beta=\left[\frac{1}{N}\sum_{j=1}^N(\frac{1}{m}{\bf X}_{j}^T{\bf X}_{j}+\lambda_j I_p)\right]^{-1}
\frac{1}{N}\sum_{j=1}^N\frac{1}{m}{\bf X}_{j}^T{\bf X}_{j}\widehat\beta_j \end{eqnarray} (see, e.g., Lin and Xi, 2011) has the bias of order $O(m^{-\delta})$ as well.
Under Condition {\it C0}, $\sqrt n\,\widehat\beta$ has a non-ignorable bias of order $O(n^{1/2-\delta}N^{\delta})$, specifically,
\begin{eqnarray}\label{(2.7)}\sqrt n Bias(\widehat\beta)=O(n^{1/2-\delta}N^{\delta})\rightarrow\infty \mbox{ if } m^{2\delta-1}=o(N).\end{eqnarray}
Therefore, the global estimator $\widehat\beta$ cannot achieve $\sqrt n$-consistency when $m^{2\delta-1}=o(N)$.


There are other examples of biased estimators (e.g., quantile estimator) satisfying that the resulting global estimators by naive average or the original DC expression have the non-ignorable bias as in (\ref{(2.4)}) and (\ref{(2.7)}).
These examples indicate that the naive average and the original DC expression are invalid when the local estimators have a non-ignorable bias. As shown in Introduction, although
bias-correction methods have been considered in the existing DC literature, the related algorithms are computationally complex, and the resulting bias-correction for global estimation is not sufficient. The observation motivates us to develop new DC methodologies.

\setcounter{equation}{0}
\section{Global bias-correction estimate in linear model}

\subsection{Modeling}

We use $\theta$ to denote the parameter vectors $\beta_S$ and $\beta$ respectively in Example 1 and Example 2, or a general parameter vector in a linear model. For convenience of modeling, suppose the dimension $p$ is fixed. The composite method proposed blow still applies to the case where $p$ depends on $n$. From the above motivating examples, we have an interesting finding: the closed representations (\ref{(2.2)}) and (\ref{(2.5)}) respectively for LASSO estimator and Ridge estimator can be expressed as the following unified form:
\begin{equation}\label{(3.1)}\widehat \theta_j=\theta+
V_{m}({\cal D}_j)\xi(\theta)+\bm\epsilon_j,\ j=1\cdots,N.\end{equation} In the above model, the matrices $V_{m}({\cal D}_j)$ depend on subsets ${\cal D}_j$, the vector $\xi(\theta)$ is a function of $\theta$, and vectors
$\bm\epsilon_j$ have zero mean. In the motivating examples, the covariance matrix $Cov[\bm\epsilon_j|{\cal D}_{j}]$ is approximately equal to a positive definite matrix $\frac1m\Sigma$. We then suppose $Cov[\bm\epsilon_j|{\cal D}_{j}]= \frac1m\Sigma$, without loss of generality.

For the LASSO estimator in Example 1,
\begin{eqnarray*}&&V_{m}({\cal D}_j)=-\left(\frac{1}{m}{\bf X}_{jS}^T{\bf X}_{jS}\right)^{-1}\lambda_j, \ \xi(\theta)=\mbox{sgn}(\theta),E[\bm\epsilon_j|{\bf X}_{jS}]=0,\\&&Cor[\bm\epsilon_j|{\bf X}_{jS}]=
\sigma^2_\varepsilon\left(\frac{1}{m}{\bf X}_{jS}^T{\bf X}_{jS}\right)^{-1}\frac{1}{m^2}{\bf X}_{jS}^T{\bf X}_{jS}\left(\frac{1}{m}{\bf X}_{jS}^T{\bf X}_{jS}\right)^{-1}\approx
\frac 1m\Sigma_S,\end{eqnarray*} where $\Sigma_S=\sigma_\varepsilon^2(E(X_SX_S^T))^{-1}$.

Similarly, for the Ridge estimator in Example 2,
\begin{eqnarray*}&&V_{m}({\cal D}_j)=-\left(\frac{1}{m}{\bf X}_{j}^T{\bf X}_{j}+\lambda_j I_p\right)^{-1}\lambda_j,\xi(\theta)=
\theta,E[\bm\epsilon_j|{\bf X}_{jS}]=0,\\&& Cov[\bm\epsilon_j|{\bf X}_{j}]
=\sigma^2_\varepsilon\left(\frac{1}{m}{\bf X}_{j}^T{\bf X}_{j}+\lambda_j I_p\right)^{-1}\frac{1}{m^2}{\bf X}_{j}^T{\bf X}_{j}\left(\frac{1}{m}{\bf X}_{j}^T{\bf X}_{j}+\lambda_j I_p\right)^{-1}\approx\frac 1m \Sigma,\end{eqnarray*}  where $\Sigma=\sigma_\varepsilon^2(E(XX^T))^{-1}$.

Let $\widehat \theta_j^k$ and $\epsilon_j^k$ be the $k$-th elements of $\widehat \theta_j$ and $\bm\epsilon_j$ respectively, and ${\bf v}^k_{m}({\cal D}_j)=V^T_{m}({\cal D}_j){\bf e}_k$, where ${\bf e}_k$ is a $p$-dimensional vector with the $k$-th element 1 and the others zero. By (\ref{(3.1)}), we have
\begin{eqnarray}\label{(3.2)}\widehat\theta_j^k=\theta^k+\xi^T(\theta){\bf v}_{m}^k({\cal D}_j)+\epsilon^k_{j}, \ j=1,\cdots,N.\end{eqnarray} Denote
$V_k=({\bf v}_{m}^k({\cal D}_1),\cdots,{\bf v}_{m}^k({\cal D}_N))^T$.
According to the motivating examples aforementioned and for constructing a valid regression, we suppose the following conditions for model (\ref{(3.2)}):
\begin{enumerate}
\item[{\it C1}.]
$E[\epsilon^k_{j}|{\cal D}_j]=0$ and $Var[\epsilon^k_{j}|{\cal D}_j]=\frac 1m\sigma^2$, where $\sigma^2$ is a positive constant.
\item[{\it C2}.] The inverse matrix
$(V_k^TV_k)^{-1}$ exists uniformly for all $N$.
\end{enumerate}
It can be seen that when $m$ is large enough, Condition {\it C1} is a direct result of the motivating examples. Thus, this condition is mild. For Condition {\it C2}, we have the following explanations.

{\it (i) The case of distribution heterogeneity.} We first consider the case where the sets ${\cal D}_1,\cdots, {\cal D}_N$ are not identically distributed. Such a distribution heterogeneity often appears under the situation of big data. A common example is streaming data, which may not be identically distributed in different observation periods. In this case, we can suppose that $V_{m}({\cal D}_1),\cdots,V_{m}({\cal D}_N)$ are not identically distributed, and consequently, the matrix $V_k^TV_k$ is invertible.

{\it (ii) The case of distribution homogeneity.} Consider the case of $X_1,\cdots,X_n$ being identically distributed observations of $X$. Under such a situation, however, Condition {\it C2} is not satisfied. To verify the point of view, we look at the LASSO estimator, in which
$V_{m}({\cal D}_j)=-\left(\frac{1}{m}{\bf X}_{jS}^T{\bf X}_{jS}\right)^{-1}\lambda_j$. When ${\cal D}_1,\cdots,{\cal D}_N$ are identically distributed and $m$ is large enough, we have $$\frac{1}{m}{\bf X}_{1S}^T{\bf X}_{1S}\approx E(X_SX_S^T),\cdots,\frac{1}{m}{\bf X}_{NS}^T{\bf X}_{NS}\approx  E(X_SX_S^T).$$
This shows that the vectors ${\bf v}_{m}^k({\cal D}_1),\cdots,{\bf v}_{m}^k({\cal D}_N)$ are approximately equal, implying that the matrix $V_k^TV_k$ is nearly degenerated, as a result, Condition {\it C2} cannot be satisfied.
We use the following method to deal with the problem.
From model (\ref{(2.1)}), we have
\begin{equation}\label{(3.3)}W_i=U_i^T\beta+a_{ij}\varepsilon_i, \ i=1,2,\cdots,m, j=1,2,\cdots, N,\end{equation} where  $W_i=a_{ij}Y_i$ and $U_i=a_{ij}X_i$ for $i\in {\cal H}_j$, and the random variables $a_{ij}$ satisfy that $a_{ij},i=1,\cdots,m$, are identically distributed for each $j$, but for $j\neq k$ $\{a_{ij},i=1,\cdots,m\}$ and $\{a_{ik},i=1,\cdots,m\}$ are not identically distributed. We then use the new variables $W_i$ and $U_i$ to construct the estimator of $\beta$.
When $m$ is large enough, we have
\begin{eqnarray*}\frac{1}{m}{\bf U}_{1}^T{\bf U}_{1}\approx  A_1E[XX^T]A_1,\cdots,\frac{1}{m}{\bf U}_{N}^T{\bf U}_{N}\approx  A_NE[XX^T]A_N,\end{eqnarray*} where $A_j={\rm diag}(a_{ij},i\in {\mathcal H}_j)$. Then, we can verify by the result above that model (\ref{(3.3)}) satisfies Condition {\it C2}. We can employ some other methods to reconstruct model (\ref{(2.1)}) such that the reconstructed model consists of non-identically distributed variables; for the details see the part of simulation studies.

Thus, both Condition {\it C1} and Condition {\it C2} can be easily satisfied. Under the two conditions, model (\ref{(3.2)}) (or (\ref{(3.1)})) could be regarded as a linear regression model, in which $\widehat\theta_j^k$ (or vector $\widehat\theta_j$) are the response variables (or response vectors), vector ${\bf v}_{m}^k({\cal D}_j)$ (or matrix $V_{m}({\cal D}_j)$) are the covariate vector (or covariate matrix), $\xi(\theta)$ is the regression coefficient, $\theta^k$ (or $\theta$) is the intercept, and $\epsilon^k_{j}$ (or $\epsilon_{j}$) are the errors. Thus, the intercept $\theta^k$ (or $\theta$) is the parameter of interest. Furthermore, models (\ref{(3.1)}) and (\ref{(3.2)}) are of DC expressions of regression. Such a structure is different from the composition methods in Lin et al. (2018), Cheng et al. (2018), Lin and Li (2008), Wang and Lin (2015), and Tong and Wang (2005). This is because these methods do not have DC structure and use a model-independent parameter (e.g., quantile and bandwidth) as an artificial covariate, which does not exist in the original model, but is identified from the estimation procedure. Moreover, these methods cannot be employed directly to the models of big data.

\subsection{Estimation}

The above modeling procedures indicate that we can apply the DC expressed model (\ref{(3.1)}) or (\ref{(3.2)}) to construct a global estimator. The use of the DC expressed model, instead of DC expressed estimation, is our main innovation. For simplicity, we mainly focus on model (\ref{(3.2)}), which has univariate ``response" $\widehat\theta_j^k$.
Under the pro forma linear regression (\ref{(3.2)}), the composite global estimator of $\theta^k$ is naturally defined as the first component of the following least squares solution:
\begin{equation}\label{(3.4)}(\widetilde\theta^k,\widetilde\xi^T)^T
=\arg\min_{\theta,\xi }\sum_{j=1}^N\left(\widehat\theta_j^k-\theta^k-\xi^T(\theta){\bf v}_{m}^k({\cal D}_j)\right)^2.\end{equation}
It can be easily verified that the composite global estimator in (\ref{(3.4)}) has the following simple expression:
\begin{equation}\label{(3.5)}\widetilde\theta^k=\overline{\widehat\theta^k}-\widetilde\xi^T\,
\overline{{\bf v}^k},\end{equation}
where $\overline{\widehat\theta^k}=\frac{1}{N}\sum_{j=1}^N\widehat\theta^k_j$, $\overline{{\bf v}^k}=\frac{1}{N}\sum_{j=1}^N{\bf v}_{m}^k({\cal D}_j)$ and
\begin{equation*}\widetilde\xi=
\left(\sum_{j=1}^N \left({\bf v}_{m}^k({\cal D}_j)-\overline{{\bf v}^k}\right)
\left({\bf v}_{m}^k({\cal D}_j)-\overline{{\bf v}^k}\right)^T\right)^{-1}
\sum_{j=1}^N \left({\bf v}_{m}^k({\cal D}_j)-\overline{{\bf v}^k}\right)
\widehat\theta^k_j.\end{equation*}

The composite global estimator is a DC expression, without accessing the raw data. The global estimator is computational simple as it is computed directly on ${\bf v}_{m}^k({\cal D}_j)$ and $\widehat\theta^k_j$, without use of any iterative algorithm and local bias-correction, and has the form of least squares.
Because of such a structure, we can construct its online updating version and make statistical inference in the case of streams (see, e.g., Schifano et al., 2016).
Furthermore, the global estimator is unbiased (see Lemma 3.2 below), because such a DC expression sufficiently uses the structural information of regression (\ref{(3.2)}) such that the unbiasedness can be achieved.
We thus call it bias-corrected global estimator (BC-GE, for short). This is totally different from the original DC expressions (see the DC expressions of the LASSO and Ridge estimators given in Subsection 2.2).

The BC-GE $\widetilde\theta^k$ in (\ref{(3.5)}) is derived from the general model framework in (\ref{(3.2)}).
Particularly, for the LASSO estimator in Example 1, the local estimators of the significant subset $\beta_S$ of $\beta$ may be different using different subsets ${\cal D}_j$. We thus employ the majority voting methods proposed by Meinshausen and Buhlmann (2010), Shah and Samworth (2013), and Chen and Xie (2014) to determine the significant subset $\beta_S$. After the significant subset $\beta_S$ is determined,
the corresponding BC-GE of the $k$-component of $\beta_S$ is
\begin{equation}\label{(3.6)}\widetilde\beta_{S}^k=\overline{\widehat\beta_S^k}-\widetilde\xi^T\,
\overline{{\bf v}_S^k},\end{equation}
where $\overline{\widehat\beta^k_S}=\frac{1}{N}\sum_{j=1}^N\widehat\beta^k_{jS}$, $\overline{{\bf v}_S^k}=\frac{1}{N}\sum_{j=1}^N{\bf v}_{m}^k({\cal D}_{jS})$,
${\bf v}_{m}^k({\cal D}_{jS})=-\left(\frac{1}{m}{\bf X}_{jS}^T{\bf X}_{jS}\right)^{-1}\lambda_j{\bf e}_k$, and
\begin{eqnarray*}&&\widetilde\xi=
\left(\sum_{j=1}^N \left({\bf v}_{m}^k({\cal D}_{jS})-\overline{{\bf v}_S^k}\right)
\left({\bf v}_{m}^k({\cal D}_{jS})-\overline{{\bf v}_S^k}\right)^T\right)^{-1}
\sum_{j=1}^N \left({\bf v}_{m}^k({\cal D}_{jS})-\overline{{\bf v}_S^k}\right)
\widehat\beta^k_j.
\end{eqnarray*}

Similarly, for the Ridge estimator in Example 2, the corresponding BC-GE is
\begin{equation}\label{(3.7)}\widetilde\beta^k=\overline{\widehat\beta^k}-\widetilde\xi^T\,
\overline{{\bf v}^k},\end{equation}
where $\overline{\widehat\beta^k}=\frac{1}{N}\sum_{j=1}^N\widehat\beta^k_{j}$,
$\overline{{\bf v}^k}=\frac{1}{N}\sum_{j=1}^N{\bf v}_{m}^k({\cal D}_j)$, and
\begin{eqnarray*}&&{\bf v}_{m}^k({\cal D}_j)=-\left(\frac{1}{m}{\bf X}_{j}^T{\bf X}_{j}+\lambda_j I_p\right)^{-1}\lambda_j{\bf e}_k,\\&&\widetilde\xi=
\left(\sum_{j=1}^N \left({\bf v}_{m}^k({\cal D}_j)-\overline{{\bf v}^k}\right)
\left({\bf v}_{m}^k({\cal D}_j)-\overline{{\bf v}^k}\right)^T\right)^{-1}
\sum_{j=1}^N \left({\bf v}_{m}^k({\cal D}_j)-\overline{{\bf v}^k}\right)
\widehat\beta^k_j.\end{eqnarray*}

\subsection{Theoretical property}

Actually, the BC-GE $\widetilde\theta^k$ given in (\ref{(3.6)}) is original least squares estimator under linear regression model (\ref{(3.2)}). Thus, its theoretical property is very simple. The following lemma follows directly from the property of the least squares estimation.

\vspace{1ex}

\noindent{\bf Lemma 3.1.} {\it Under Conditions {\it C1} and {\it C2}, the BC-GE $\widetilde\theta^k$ given in (\ref{(3.6)}) has mean and variance as
$$E[\widetilde\theta^k|V_k]=\theta^k, \ Var[\widetilde\theta^k|V_k]=\frac 1m\sigma^2{\bf e}^T_1\left(({\bf 1},V_k)^T({\bf 1},V_k)\right)^{-1}{\bf e}_1.$$
}

According to the two motivating examples, we have ${\bf v}_{m}^k({\cal D}_j)=O(\lambda_j)$, which tend to zero as $m\rightarrow\infty$.
Note that the sizes of all the subsets ${\cal D}_j$ are supposed to be identical. Thus, we assume $\lambda_j=\lambda$ for all $ j$ in the subsection, without loss of generality. As a result, we have the following condition:
\begin{enumerate}
\item[{\it C3}.] ${\bf v}_{m}^k({\cal D}_j)=O(\lambda)$ for $j=1,\cdots,m$. \end{enumerate} Then,
Conditions {\it C1} - {\it C3}, and Lemma 3.1 together result in the following lemma.

\noindent{\bf Lemma 3.2.} {\it Under Conditions {\it C1} - {\it C3}, the variance of the BC-GE $\widetilde\theta^k$ satisfies $$Var(\widetilde\theta^k|V_k)=O_p(n^{-1}).$$}

Consequently, we have the following main results.

\vspace{1ex}

\noindent{\bf Theorem 3.3.} {\it Under Conditions {\it C1} - {\it C3}, the BC-GE $\widetilde\theta^k$ is always $\sqrt n$-consistent for arbitrary choices of $N$ and $m$.}

The theorem guarantees the standard consistency rate for any choices of $N$ and $m$.
Such a result cannot be attained by the existing methods.
Furthermore, in order to establish the asymptotic normality, we need the condition:
\begin{enumerate}
\item[{\it C4}.] The following limits exist:
\begin{eqnarray*}\frac{1}{N\lambda}\sum_{j=1}^N{\bf v}_{m}^k({\cal D}_j)\stackrel{p}\rightarrow E[{\bf v}^k],\ \frac{1}{N\lambda^2}\sum_{j=1}^N{\bf v}_{m}^k({\cal D}_j)({\bf v}_{m}^k({\cal D}_j))^T\stackrel{p}\rightarrow E[{\bf v}^k({\bf v}^k)^T].\end{eqnarray*}
\end{enumerate} This condition comes from {\it C3} and the motivating examples.
In the above, the notation $E[{\bf v}^k]$ stands for a fixed number, but is not always the expectation of a vector ${\bf v}^k$, and the notation $E[{\bf v}^k({\bf v}^k)^T]$ denotes a fixed matrix, but is not always the expectation of a matrix ${\bf v}^k({\bf v}^k)^T$. This is because ${\bf v}_{m}^k({\cal D}_j),j=1,\cdots,N$, may not be random, and even for the case of random variables, they may not be identically distributed.
Obviously, the above condition is common. With this condition, the asymptotic normality holds; the following theorem states the details.

\noindent{\bf Theorem 3.4.} {\it Under Conditions {\it C1} - {\it C4}, the BC-GE $\widetilde\theta^k$ has the asymptotic normality as
$$\sqrt n\left(\widetilde\theta^k-\theta^k\right)\stackrel{d}\rightarrow N\left(0,\sigma^2 \left(1+E[({\bf v}^k)^T]\left(Cov[{\bf v}^k]\right)^{-1}E[{\bf v}^k]\right)\right)\ (n\rightarrow\infty) $$ for any choices of $N$ and $m$.
 }

By the theorem, we can compare the BC-GE with the full data estimator that is supposed to be computed on the entire data set. Theorem 3.4 and the unbiasedness in Lemma 3.1 imply that the mean square error of the BC-GE is usually larger than that of the unbiased full data estimator (e.g., the full data least squares estimator for linear regression model). However, if the full data estimator is biased, the improvement of the BC-GE is significant. In the following, we use the full data LASSO estimator as an example to illustrate this point of view. Let $\beta^k$ be the $k$-th component of $\beta_S$ as in Example 1. Then, the full data LASSO estimator $\widehat\beta^k$ has the mean square error as
\begin{eqnarray}\label{(3.8)}\nonumber&&
MSE[\widehat\beta^k]\\\nonumber&&=(\mbox{sgn}(\beta_S))^T E[{\bf v}_n^k](E[{\bf v}_n^k])^T\mbox{sgn}(\beta_S)+(\mbox{sgn}(\beta_S))^TCov[{\bf v}_n^k]\mbox{sgn}(\beta_S)+\frac{\sigma^2}{n}\\&&=(\mbox{sgn}(\beta_S))^T E[{\bf v}_n^k({\bf v}_n^k)^T]\mbox{sgn}(\beta_S)+\frac{\sigma^2}{n},\end{eqnarray} where ${\bf v}_n^k=-\left(\frac{1}{n}{\bf X}_S^T{\bf X}_S\right)^{-1}\lambda{\bf e}_k$, and  $\lambda=O(n^{-\delta})$ for some constant $0<\delta<1$.
The proof of (\ref{(3.8)}) is given in Appendix. When the full data LASSO estimator has a non-ignorable bias (i.e., $0<\delta<1/2$), the BC-GE $\widetilde\beta^k$ is much better than the full data LASSO estimator $\widehat\beta^k$ because $\sqrt n\,\widetilde\beta^k$ has an finite MSE, while the MSE of $\sqrt n\,\widehat\beta^k$ tends to infinity. When $\delta=1/2$ (i.e., $\lambda=cn^{-1/2}$ for a constant $c>0$), then
\begin{equation*}
MSE[\sqrt n\,\widehat\beta^k]=c^2(\mbox{sgn}(\beta_S))^T E[{\bf v}_n^k({\bf v}_n^k)^T]\mbox{sgn}(\beta_S)+\sigma^2.\end{equation*} It shows that when ${\bf v}^k_{m}({\cal D}_j),j=1,\cdots,N$, are very dispersed, $MSE[\sqrt n\,\widehat\beta^k]$ is larger than $MSE[\sqrt n\,\widetilde\beta^k]$. In this case, the BC-GE is better than the full data LASSO estimator as well. If $1/2<\delta<1$, however, $MSE[\sqrt n\,\widehat\beta^k]$ is smaller than $MSE[\sqrt n\,\widetilde\beta^k]$ when $n$ is large enough.

All the theoretical properties aforementioned will be illustrated by the simulation studies given in Section 5.

\setcounter{equation}{0}
\section{Extensions}

We extend the method proposed above into the cases of nonlinear and nonparametric models.

\subsection{Global bias-correction estimate in nonlinear model}

Consider the following nonlinear model: \begin{eqnarray}\label{(4.1)}Y_i=q(\theta,X_i)+\varepsilon_i,i=1,\cdots,n,\end{eqnarray} where $q(\cdot,\cdot)$ is a given function, and the error term satisfies $E(\varepsilon|X)=0$ and
$Var(\varepsilon|X)=\sigma^2_\varepsilon$. The parameter $\theta$ can be estimated, for example, by least squares method. More generally, we consider the following $M$- and $Z$-estimators of $\theta$. For the case of $M$-estimator, the local estimator $\widehat\theta_j = g_j({\cal D}_j)$ is defined as the minimizer of the following objective function: $$M_{j}(\theta)=\frac{1}{m}\sum_{i\in{\cal H}_j}m(\theta,Z_i),$$ where $m(\theta,z)$ is a given function. A common choice of $m(\theta,z)$ is $(y-q(\theta,x))^2$, which corresponds to leat squares estimator. For the case of $Z$-estimator, the local estimator $\widehat\theta_j = g_j({\cal D}_j)$ is defined as the solution of the following equation:
$$\Psi_{j}(\theta)=\frac{1}{m}\sum_{i\in{\cal H}_j}\psi(\theta,Z_i)=0,$$ where the estimating function $\psi(\theta,z)=(\psi_1(\theta,z),\cdots,\psi_p(\theta,z))^T$ is a known $p$-dimensional vector-valued function satisfying $E[\psi(\theta,Z)]=0$. For example, $\psi(\theta,z)$ can be chosen as the derivative of $m(\theta,z)$ with respect to $\theta$ if it exists. Under some regularity conditions (see, e.g., van der Vaart, 1998; Jure\v{c}kov\'{a}, 1985;  Jure\v{c}kov\'{a} and Sen, 1987), we have the following asymptotic representation:
\begin{equation*}\widehat \theta_j=\theta-
\frac{1}{\sqrt m}D^{-1}(\theta)\frac{1}{\sqrt m}\sum_{i\in{\cal H}_j}\psi(\theta,Z_i)+O_p\Big(\frac{1}{m^\gamma}\Big),\ \ j=1\cdots,N,\end{equation*} where $D(\theta)$ is the derivative matrix of $E[\psi(\theta,z)]$ with respective to $\theta$ if it exists, and $\gamma$ is a constant satisfying $1/2<\gamma\leq 1$. It
is known that $\gamma =1$ if $\psi(\theta,z)$ is twice differentiable with respect to $\theta$, but $\gamma=3/4$ if $\psi(\theta,z)$ has jump discontinuities; see, for example, Jure\v{c}kov\'{a} (1985), Jure\v{c}kov\'{a} and Sen (1987), and He and Shao (1996). By the two methods, the local estimator is biased usually. Moreover, according to Bontemps (2018), we suppose that $\psi(\theta,Z)$ is a robust moment condition in the sense of
$$\frac{1}{\sqrt m}\sum_{i\in{\cal H}_j}\psi(\theta,Z_i)=\frac{1}{\sqrt m}\sum_{i\in{\cal H}_j}\psi(\widehat\theta^0,Z_i)+o_p(1),$$ where $\widehat\theta^0$ is an initial estimator of $\theta$ computed on a subset. Consequently,
\begin{equation}\label{(4.2)}\widehat \theta_j=\theta-
\frac{1}{\sqrt m}D^{-1}(\theta)\frac{1}{\sqrt m}\sum_{i\in{\cal H}_j}\psi(\widehat\theta^0,Z_i)+o_p\Big(\frac{1}{\sqrt m}\Big),\ \ j=1\cdots,N.\end{equation}
By (\ref{(4.2)}) and the same argument as used in (\ref{(3.2)}), we get the following pro forma linear model:
\begin{eqnarray}\label{(4.3)}\widehat\theta_j^k=\theta^k+\xi^T(\theta)\psi({\cal D}_j)+\epsilon_{j}, \ j=1,\cdots,N,\end{eqnarray} where $\xi^T(\theta)=-
{\bf e}_k^TD^{-1}(\theta)$ and $\psi({\cal D}_j)=\frac{1}{m}\sum_{i\in{\cal H}_j}\psi(\widehat\theta^0,Z_i)$. The main difference from model (\ref{(3.2)}) is that here the error $\epsilon_{j}$ is not  unbiased for zero. Actually, it is an infinitesimal of higher order than $\xi^T(\theta)\psi({\cal D}_j)$. Then, by the above model and the same argument as used in (\ref{(3.6)}), we get the BC-GE of $\theta^k$ as
\begin{equation}\label{(4.4)}\widetilde\theta^k=\overline{\widehat\theta^k}-\widetilde\xi^T\,
\overline{\psi},\end{equation}
where $\overline{\widehat\theta^k}=\frac{1}{N}\sum_{j=1}^N\widehat\theta^k_j$, $\overline{\psi}=\frac{1}{N}\sum_{j=1}^N\psi({\cal D}_j)$ and
\begin{equation*}\widetilde\xi=
\left(\sum_{j=1}^N \left(\psi({\cal D}_j)-\overline{\psi}\right)
\left(\psi({\cal D}_j)-\overline{\psi}\right)^T\right)^{-1}
\sum_{j=1}^N \left(\psi({\cal D}_j)-\overline{\psi}\right)
\widehat\theta^k_j.\end{equation*}

The key for a valid estimator is that the matrix $\sum_{j=1}^N \left(\psi({\cal D}_j)-\overline{\psi}\right)
\left(\psi({\cal D}_j)-\overline{\psi}\right)^T$ is invertible. We thus need the condition: the model is fixed design, or the data sets ${\cal D}_j,j=1,\cdots,N$, are not identically distributed, or the data sets $\psi({\cal D}_j),j=1,\cdots,N$, are transformed such that the resulting data sets are not identically distributed; for more details see the related discussions in Subsection 3.1.

Because the expectation of $\epsilon_j$ is not zero and $\psi({\cal D}_j)$ depends on the initial estimator $\widehat\theta^0$, the theoretical property of the BC-GE $\widetilde\theta^k$ in (\ref{(4.4)}) is different from or more complex than those in linear model. Furthermore, when  $\psi(\theta,Z)$ does not satisfy the robust moment condition, the difference between $\frac{1}{\sqrt m}\sum_{i\in{\cal H}_j}\psi(\theta,Z_i)$ and $\frac{1}{\sqrt m}\sum_{i\in{\cal H}_j}\psi(\widehat\theta,Z_i)$ is non-ignorable. In this case, we cannot construct a pro forma linear regression as in (\ref{(4.3)}). These issues will be investigated in the future.

\subsection{Global bias-correction estimate in nonparametric model}

 Consider the following
nonparametric regression:
$$Y_i=r(X_i)+\varepsilon_i,i=1,\cdots,n,$$ where $r(x)$ is a smooth nonparametric regression function for
$x\in [0,1]$, and the error term satisfies $E(\varepsilon|X)=0$ and
$Var(\varepsilon|X)=\sigma_\varepsilon^2$.
Under certain regularity conditions
(see, e.g., Bhattacharya and Gangopadhyay, 1990;
Chaudhuri, 1991; Hong, 2003), a commonly used kernel estimator $\widehat r_j(x)$ (e.g., N-W estimator) computed on ${\cal D}_j$ has following Bahadur representation:
\begin{eqnarray}\label{(4.5)}\widehat r_j(x)=r(x)+v^{-1}_h(x) \frac{1}{m}\sum_{i\in{\cal H}_j}K_{h}(X_i-x)(Y_i-r(x))+O_p\left(\frac{1}{m^{3(1-\varsigma)/4}}\right)
\end{eqnarray} for $x\in[0,1]$ and $j=1,\cdots,N$,
where $K_h(x)=h^{-1}K(x/h)$, $K(\cdot)$ is a kernel function, $h$ is bandwidth satisfying $h=O(m^{-\varsigma})$ for some
constant $0<\varsigma<1$, and $v_h(x)=E[K_h(X-x)]$. Suppose $m=n^\tau$ for some constant $\tau$ satisfying $\varsigma<1-2\tau/3$. Then, the error term $O_p\left(1/m^{3(1-\varsigma)/4}\right)$ is an infinitesimal of higher order than the second term on the right hand side of (\ref{(4.5)}). In this case the local estimator is always biased.

By (\ref{(4.5)}) and the same argument as used in Subsection 4.1, we get the following pro forma linear model:
\begin{eqnarray}\label{(4.6)}\widehat r_j(x)=r(x)+\alpha(x)\phi_h(x,{\cal D}_j)+\epsilon_j,j=1,\cdots,N,
\end{eqnarray} where $\alpha(x)=v^{-1}_h(x)$ and $\phi_h(x,{\cal D}_j)=
\frac{1}{m}\sum_{i\in{\cal H}_j}K_{h}(X_i-x)(Y_i-\widehat r^0(x))$, and $\widehat r^0(x)$ is an initial estimator of $r(x)$ computed on a subset. Then, by the above model and the same argument as used previously, we get the BC-GE of $r(x)$ as
\begin{equation}\label{(4.7)}\widetilde r(x)=\overline{r}(x)-\widetilde\alpha_h(x)\,
\overline{\phi}_h(x),\end{equation}
where $\overline{r}(x)=\frac{1}{N}\sum_{j=1}^N\widehat r_j(x)$, $\overline{\phi}_h(x)=\frac{1}{N}\sum_{j=1}^N\phi_h(x,{\cal D}_j)$ and
\begin{eqnarray*}\widetilde\alpha_h(x)&=&
\left(\sum_{j=1}^N \left(\phi_h(x,{\cal D}_j)-\overline{\phi}_h(x)\right)
\left(\phi_h(x,{\cal D}_j)-\overline{\phi}_h(x)\right)^T\right)^{-1}\\&&\times
\sum_{j=1}^N \left(\phi_h(x,{\cal D}_j)-\overline{\phi}_h(x)\right)
\widehat r_j(x).\end{eqnarray*}
Because of the nonzero expectation of $\epsilon_j$, the dependence between the estimator and the choice of $h$ and the correlation between $\phi_h(x,{\cal D}_j)$ and the estimator $\widehat r_j(x)$, the theoretical property of the BC-GE $\widetilde r(x)$ in (\ref{(4.7)}) is more complex than those aforementioned. Moreover, similar to the case of nonlinear regression aforementioned, when the second term on the right hand side of (\ref{(4.5)}) is not a robust moment condition in the sense of Bontemps (2018), the error term in (\ref{(4.6)}) is non-ignorable. These issues will be investigated in the future as well.

\setcounter{equation}{0}
\section{Simulation Studies}

The goal of this section is to comprehensively evaluate the performance of the proposed method by a series of simulations. To this end, the newly proposed BC-GE for biased LASSO and Ridge and N-W estimators is compared respectively with the naive averaging estimators and DC-expression estimators (\ref{(2.3)}) and (\ref{(2.6)}) from LASSO and Ridge estimators in linear model, and the naive averaging estimators from N-W estimator in nonparametric model. Various experiment conditions such as the correlation among data, and heterogeneity or homogeneity of the distributions of data are overall considered in the procedures of simulation studies.
As an object of reference, the full data estimator that is computed on the entire dataset is considered as well. The mean squared error (for the parametric
model) and the mean integrated squared error (for the nonparametric model)
are used to measure the performance of the involved estimators. The simulation results of the estimation bias are also reported for checking the bias-correction of the new method. All the criterions computed are based on 500 repetitions.

\subsection{Linear model with heterogeneously distributed data}

{\it Experiment 1. LASSO-based estimators.}
Here we investigate the performance of the BC-GE for biased LASSO estimator. Reference to Chen et al. (2018) and Battey et al. (2108), the dataset with size $n=10000$ are generated from the linear model
\bea
Y=X^T\beta+\varepsilon,
\eea
where $\beta=(3,2,1,-2,0,0,\cdots,0)^T$, a 20-dimensional vector, and $\varepsilon$ follows the standard normal distribution $N(0,1)$. In the procedure of simulation, the heterogeneously distributed data $X_i\in \mathcal D_j$ are generated from $N_p(\mu_{j},\Sigma)$, where $\Sigma=(\sigma_{kl})_{p\times p}$ with $\sigma_{kl}=0.5^{|k-l|}$, and $\mu_{j}$ are generated from $N_p(0,I_p)$.
The number of batches $N$ takes the values 10, 20, 50, 100 and 200, respectively.
For the linear model above, we mainly focus on the significant subset $\beta_S$ of $\beta$, i.e., $\beta_S=(\beta^{k_1},\cdots,\beta^{k_s})^T$ with $\beta^{k_t}\neq 0$ for $t=1,\cdots,s$. As shown in Subsection 3.2, the local estimators of $\beta_S$ may be different across different subsets ${\cal D}_j$, thus, the majority voting method is employed to determine the significant subset $\beta_S$. The penalty parameters $\lambda_j, j=1,2,\cdots, N$, are selected by 5-fold cross-validation. For the details see Meinshausen and Buhlmann (2010), Shah and Samworth (2013), and Chen and Xie (2014).

\begin{center}(Figure 1 and Figure 2 about here)\end{center}

Figure 1 shows the empirical bias of all the estimators considered, and Figure 2 presents the estimated mean square error of the involved estimators. We have the following findings:
\begin{enumerate}
\item[1)] The newly proposed BC-GE performs comparably well with the full data estimator. Actually, the difference between the BC-GE and full data estimator is negligible, and the bias and mean square error of both estimators are nearly zero for any choices of $N$.
\item[2)] Under criteria of estimation bias and mean square error, the BC-GE is much better than the naive averaging estimator and the DC-expression estimator uniformly for any choices of $N$. Furthermore, the bias and mean square error of the naive averaging estimator and DC-expression estimator are increasing with the number $N$, and both estimators are almost collapsed when $N$ is large.
\item[3)] The naive averaging estimator is the worst one among the estimators considered for any choices of $N$.\end{enumerate}

\noindent
{\it Experiment 2. Ridge-based estimators.}
Here we examine the behavior of the BC-GE for the Ridge estimation. For the linear regression model, the regression coefficients are chosen as $\beta=(2,0.5,-1,-2)^T$, a 4-dimensional vector, and the covariance matrix of the covariate vector $X$ is chosen as $\Sigma=(\sigma_{ij})_{4\times 4}$ with $\sigma_{ij}=0.99^{|i-j|}$. The other experiment conditions are designed as the same as those in Experiment 1. Because this is non-sparse and low dimensional regression, and the correlation among the components of $X$ is relatively strong, we can use the Ridge estimation method to estimate $\beta$.

\begin{center}(Figure 3 and Figure 4 about here)\end{center}

Figure 3 and Figure 4 report the empirical bias and mean square error of all the estimators. It can be seen that the fashions of the simulation results in Figure 3 and Figure 4 are the almost same as those in Figure 1 and Figure 2 of Experiment 1. In brief, the BC-GE is the best one, the naive averaging estimator is the worst one among all the estimators for any choices of $N$, and particularly, when $N$ is large, the BC-GE is significantly better than the naive averaging estimator and the DC-expression estimator.

%

\subsection{Linear model with identically distributed data }
{\it Experiment 3. LASSO-based estimators.}
The model settings are the same with those in Experiment 1, except for that the predictors $X_i$ in each batch $\mathcal D_j$ are all generated from distribution $N_p(0, \Sigma)$, i.e., the only difference between this experiment and Experiment 1 is that the data in this experiment are homogeneously distributed, but the data in Experiment 1 are heterogeneously distributed. To guarantee the Condition {\it C2}, data ${\bf X}_j$ and ${\bf Y}_j$ in batch $\mathcal D_j$ are both multiplied by matrix $A_j$, where $A_j={\rm diag}(a_{j1},a_{j2},\cdots,a_{jm})$, $a_{jk}, k=1,2,\cdots, m$, are generated from normal distribution $N(\mu_j,1)$, $\mu_j=1+9(j-1)/N$.

\begin{center}(Figure 5 and Figure 6 about here)\end{center}

Figure 5 and Figure 6 present the bias and mean square error of all the estimators. Similar to the case of identically distributed data, the BC-GE is the best one among all the estimators for any choices of $N$, which has the similar behavior to that of the full data estimator. Particularly, when $N$ is large, the BC-GE is significantly better than the naive averaging estimator and the DC-expression estimator.

\

\noindent
{\it Experiment 4. Ridge-based estimators.}
The model settings are the same as those in Experiment 2, except for that the data $X_j$ in each batch are all generated from a common population, $X\sim N_p(0,\Sigma)$, and the regression coefficients are set as $\beta=(3,2,-1,-2)^T$. To guarantee the Condition {\it C2}, the similar strategies as in Experiment 3 are employed to generate heterogenous data. The GCV criterion is employed to choose the penalty parameters $\lambda_j, j=1,2,\cdots, N$.

\begin{center}(Figure 7 and Figure 8 about here)\end{center}

Figure 7 and Figure 8 show the bias and mean square error of all the estimators. As can be seen from the figures, the proposed global estimator performs comparably well with the estimator based on the full data, moreover, it behaves significantly well in bias reduction for the ridge estimator, while the naive estimator performs worst among the four estimators.

\subsection{Nonparametric model }

Finally, we briefly examine the behavior of the new method in nonparametric model, although in the case the method has not been completely clarified and the related theoretical property has not been investigated aforementioned in Section 4.

\noindent
{\it Experiment 5. N-W-based estimators.} Consider the following nonparametric regression
$$Y_i=r(X_i)+\varepsilon_i, \ i=1,\cdots,n,$$ where $X_i\sim U(0,1)$, the errors are chosen as $\varepsilon_i\sim N(0,0.5^2)$, the regression function is designed as $r(x)=\sin(2\pi x)+2\exp(x^2)$ and the
sample size $n$ takes value 10000. The entire dataset are divided into $N (N=10,20,50,100,200)$ batches with equal size $m=n/N$.

In this experiment,
the Gaussian kernel $K(u)=1/\sqrt{2\pi}\exp\{-u^2/2\}$ is
employed to construct kernel estimators, and cross-validation is applied to select bandwidth $h_j,j=1,2,\cdots,N$. The simulation results are reported in Table~1, where the MISE stands for the empirical mean integrated squared error through $500$ repetitions. Moreover, the quantile curves of  the BC-GE, naive averaging estimator and full data estimator for $r(x)$ are also presented. Because the results are similar for different choices of batch $N$, we only show the quantile curves for $N=50$ in Figure~9. Each subfigure contains $0.05$, $0.5$, and $0.95$ quantile curves of the nonparametric estimator and the true curve of $r(x)$.
\vspace{-1ex}
\begin{table}\small
\center
\caption{MISEs for the nonparametric regression estimators in Experiment 5}
\label{tab:1}
\vspace{0.3cm}
\begin{tabular}{ccccc}
  \hline
  Num. of Batch ($N$)  & BC-GE & Full data & Naive \\
  \hline

  10 &  2.6179 (0.7112) & 3.7971 (1.6750)  & 4.8143 (2.1532)\\
  20 &  2.9145 (0.7443) & 3.7924 (1.6744)  & 4.8426 (2.1948) \\
  50 &  3.4878 (0.8774) & 3.8146 (1.7171)  & 5.1255 (2.3524) \\
  100&  5.8981 (1.3130) & 3.6669 (1.6132)  & 6.9162 (2.9409)\\
  200&  6.5604 (1.6244) & 3.6085 (1.5526)  & 7.4841 (3.0450)\\
  \hline
  \end{tabular}

  {\footnotesize Note: MISE and its standard deviation(in parenthesis) is in the scale of $\times 10^{-4}$}
\end{table}

\begin{center}
(Figures~9 about here)\vspace{-1ex}
\end{center}
By comparing the MISEs and the quantile curves of
the three estimators in Table 1 and Figures 9, respectively, we have the following findings: (1) Usually, the BC-GE estimator works well with small MISE compared with the full data and naive estimators; (2) The naive estimator performs worst among these estimators. Unlike the case of linear model, however, the number of batch clearly affects the performance of the BC-GE and naive averaging estimator. Note our method is based on (\ref{(4.6)}) and (\ref{(4.7)}), the estimating equation is not robust in the sense of Bontemps (2018). Thus, new technique (e.g., robust estimation equation method) should be developed in the future to improve the new method.

\setcounter{equation}{0}
\section{Conclusions and future works}

In this paper, we established a global bias-correction divide-and-conquer framework for biased estimation under the case of big data. Our method for composition is based on a closed representation of the local estimators obtained by the data in each batch.
Thus, the main difference from the classical DC method is that the new GBC-DC method can absorb the information hidden in the statistical structure and the variables in each batch of data.
By such a representation and least squares, the resulting global estimator is strictly unbiased even if the local estimators have a non-negligible bias. On the other hand, the new method is simple and computationally efficient, without use of any iterative algorithm and local bias-correction. The theoretical properties show that new method behaves as the full data estimator for any choice of the number of batches. Moreover, our comprehensive
simulation studies illustrate that the proposed GBC-DC approach is significantly bias-corrected, and the behavior is comparable with the full data estimation and is much better than the competitors.

Although we mainly fucus on linear model, our method can be extended into other models such as nonlinear and  nonparametric regression models. However, some new techniques should be developed for these extensions. It is because the related representation is unprecise and contains a plug-in estimator. As a result, the theoretical property is difficult to be established and the finite sample behavior is not better than these in linear model. These are interesting issues and are worth further study in the future.

\

\

\centerline{\Large \bf References}

\baselineskip=18pt

\begin{description}


\item Battey, H., Fan, J., Liu, H., Lu, J. and Zhu, Z. (2018). Distributed testing and estimation under sparse high dimensional models. {\it Ann. Statist.}, {\bf 46}, 1352-1382.

\item Bhattacharya, P. K. and Gangopadhyay, A.
(1990). Kernel and nearest neighbor estimation of a conditional
quantile. {\it Ann. Statist.}, {\bf 18}, 1400-1415.

\item Bontemps, C. (2018). Moment-based tests under parameter uncertainty. {\it Review of Economics and Statistics} (To appear).

\item Bradic, J., Fan, J. and Wang, W. (2011).
Penalized composite quasi-likelihood for ultrahigh
dimensional variable selection. {\it J. R. Statist. Soc.}, B,
{\bf 73}, 325-349.

\item Chaudhuri, P. (1991). Nonparametric estimates of regression
quantiles and their local Bahadur representation. {\it Ann.
Statist.}, {\bf 19}, 760-777.

\item Chen, X., Liu, W. and Zhang, Y. (2018). Quantile regression under
memory constraint. {\it Ann.
Statist.} To appear.

\item Chen, X. and Xie, M. (2014). A split-and-conquer approach for analysis of extraordinarily
large data. {\it Statist. Sinica}, 1655-1684.

\item Chen, Y. X., Dong, G. Z., Han, J. W., Pei, J.,
Wah, B. W. and Wang, J. Y. (2006). Regression cubes with lossless compression and aggregation. {\it
IEEE Transaction on Knowledge and Data Engineering}, {\bf 18}, No. 12, 1-14.

\item Cheng, M. Y., Huang, T, Liu, P. and and Peng, H. (2018). Bias reduction for nonparametric and semiparametric regression models. {\it Statistica Sinica},  {\bf 28}, 2749-2770.

\item Dai, W. L., Tong, T. J. and Zhu, L. X (2017), On the choice of difference sequence in a unified framework for variance estimation in nonparametric regression. {\it Statistical Science}, {\bf 32}, 455-468.

\item Dai, W. L., Tong, T. J. and Genton, M. G. (2016), Optimal estimation of derivatives in nonparametric regression. {\it Journal of Machine Learning Research}, {\bf 17}, 1¨C25.

    \item Fan, J. and Wang, W. (2011). Penalized composite quasi-likelihood for ultrahigh
dimensional variable selection. {\it J. R. Statist. Soc}. B, 73, 325-349.

\item Greenwald, M. B. and Khanna, S. (2004). Power-conserving computation of order statistics
over sensor networks. In {\it Proceedings of the ACM Symposium on Principles of
Database Systems}.

\item Guha, S. and Mcgregor, A. (2009). Stream order and order statistics: quantile estimation
in random order streams. {\it SIAM J. Comput}., {\bf 38}, 2044-2059.

\item He, X. M. and Shao, Q. M. (1996). A general Bahadur representation of $M$-estimators and its application to linear regression with nonstochastic designs.  {\it Annals of Statistics},
{\bf 24}, 2608-2630.

\item Hong, S. Y. (2003). Bahadur representation and its
applications for local polynomial estimation in nonparametric
$M$-regression. {\it Nonparametric Statistics}, {\bf 15}, 237-251.


\item Jordan, M. I., Lee, J. D. and Yang, Y. (2018). Communication-efficient distributed
statistical inference. {\it J. Amer. Statist. Assoc}. To appear.

\item Huang, J., Horowitz, J. L. and Ma, S. (2008). Asymptotic properties of
Ridge estimation in spare high-dimensional regression models. {\it
Ann. Statist.}, {\bf 36}, 578-613.

\item Jure\v{c}kov\'{a}, J. (1985). Representation of $M$-estimators with the second-order asymptotic distribution. {\it Statist. Decisions}, {\bf 3}, 263-276.

\item Jure\v{c}kov\'{a} J. and Sen, P. K. (1987). A second-order asymptotic distributional representation of
$M$-estimators with discontinuous score functions. {\it Ann. Probab.}, {\bf 15} 814-823.

\item Kai, B, Li, R. and Zou, H. (2010). Local composite quantile regression smoothing: an efficient
and safe alterative to local polynomial regression. {\it J. R. Statist. Soc}. B, {\bf 72}, 49-69.

\item Kai, B, Li, R. and Zou, H. (2011). New efficient estimation and variable selection methods for
semiparametric varying-coefficient partially linear models. {\it
Ann. Statist.}, {\bf 39}, 305-332.

\item Knight, K. and Fu, W. (2000). Asymptotics for Lasso-type estimators. {\it Ann. Statist.}, {\bf 28}, 1356-1378.

\item Lee, J. D., Liu, Q., Sun, Y. and Taylor, J. E. (2017). Communication-efficient sparse
regression. {\it J. Mach. Learn. Res.}, {\bf 18}, 1-30.

\item Li, R., Lin, D. K. and Li, B. (2013). Statistical inference in massive data sets. {\it Appl.
Stoch. Model Bus}. {\bf 29}, 399-409.

\item Li, K. and Yang, J. (2018). Score-Matching Representative Approach for Big Data
Analysis with Generalized Linear Models. Available via http://arxiv.org/ abs/1811.00462?context=stat.

\item Lian, H., Zhao, K. and Lv, S. G. (2018). Projected spline estimation of the nonparametric function in high-dimensional partially linear models for massive data. {\it Ann. Statist.} (To appear).

\item Lin, L. and Li, F. (2008). Stable and bias-corrected estimation for nonparametric
regression models. {\it Journal of Nonparametric Statistics},
{\bf 20}, 283-303.

\item Lin, L.,  Li, F., Wang, K. N. and Zhu, L. X. (2019). Composite estimation: An asymptotically weighted least squares approach. {\it Statistica Sinica} {\bf 29}, 1367-1393.

\item Lin, N. and Xi, R. (2011). Aggregated Estimating Equation Estimation. {\it Statistics and
Its Interface}, {\bf 4}, 73-83.


\item Manku, G. S., Rajagopalan, S. and Lindsay, B. G. (1998). Approximate medians
and other quantiles in one pass and with limited memory. In {\it Proceedings of the ACM
SIGMOD International Conference on Management of Data}.

\item Mcdonald, R., Mohri, M., Silberman, N., Walker, D. and Mann, G. S. (2009).
Efficient large-scale distributed training of conditional maximum entropy models. In {\it
Advances in Neural Information Processing Systems}, 1231-1239.

\item Meinshausen, N. and Buhlmann, P. (2010). Stability selection. {\it J. Roy. Statist. Soc. Ser.}, B {\bf 72}, 417-473.



\item Jonathan, D. R., and Boaz, N. (2016). On the optimality of averaging in distributed statistical learning. {\it Information and Inference: A Journal of the IMA}, {\bf 5}, 379-404.

\item Shah, R. and Samworth, R. J. (2013). Variable selection with error control: Another look at
stability selection. {\it J. Roy. Statist. Soc. Ser}., B {\bf 75}, 55-80.

\item Schifano, E. D., Wu, J., Wang, C., Yan, J., and Chen, M. H. (2016). Online updating of statistical inference in the big data setting. {\it Technometrics}, {\bf 58} (3), 393-403.

\item Shi, C., Lu, W. and Song, R. (2017). A massive data framework for $M$-estimators with
cubic-rate. {\it J. Amer. Statist. Assoc}. To appear.


\item Sun, J., Gai, Y. J. and Lin, L. (2013). Weighted local linear composite quantile estimation for the case of general error distributions. {\it Journal of Statistical Planning and Inference}, {\bf 143}, 1049-1063.

\item Tibshirani, R. (1996). Regression shrinkage and selection via the lasso. {\it Journal
of the Royal Statistical Society}, Series B, {\bf 58}, 267-288.

\item Tong, T. and Wang, Y. (2005). Estimating residual variance in nonparametric regression using least
squares. {\it Biometrika}, {\bf 92}, 821-830.

\item van der Vaart, A. W. (1998). {\it Asymptotic Statistics}. Cambridge University
Press.

\item Volgushev, S., Chao, S.-K. and Cheng, G. (2018). Distributed inference for quantile
regression processes. {\it Ann. Statist.} To appear.

\item Wainwright, M. J. (2009). Sharp threshold for high-dimensional and noisy sparsity recovery using
$\ell_1$-constrained quadratic programming (Lasso). {\it IEEE Transactions on Information Theory}, {\bf 55}, 2183-2202.

\item  Wang, H., Yang, M. and Stufken, J. (2018) Information-based optimal
subdata selection for big data linear regression. {\it Journal
of the American Statistical Association}. To appear, available via
https://haiying-wang.uconn.edu/wp-content/uploads/sites/2127/2017/04/IBOSS\_Linear.pdf

\item Wang, J., Kolar, M., Srebro, N. and Zhang, T. (2017). Efficient distributed learning
with sparsity. In {\it Proceedings of the International Conference on Machine Learning}.

\item Wang, W. W. and Lin, L. (2015). Derivative estimation based on difference sequence via
locally weighted least squares regression. {\it Journal of Machine Learning Research}, {\bf 16},  2617-2641.

\item Wang. W. W., Yu, P., Lin, L. and Tong, T. J. (2019). Robust Estimation of Derivatives Using Locally Weighted Least Absolute Deviation Regression. {\it Journal of Machine Learning Research}, (to appear).

\item Zinkevich, M., Weimer, M., Li, L., and Smola, A. J. (2010). Parallelized stochastic
gradient descent. In {\it Advances in Neural Information Processing Systems}, 2595-2603.

\item Zhang, Y., Duchi, J. and Wainwright, M. (2015). Divide and conquer kernel ridge
regression: A distributed algorithm with minimax optimal rates. {\it J. Mach. Learn. Res.},
{\bf 16}, 3299-3340.

\item Zhang, Q. and Wang, W. (2007). A fast algorithm for approximate quantiles in high
speed data streams. In {\it Proceedings of the International Conference on Scientific and
Statistical Database Management}.

\item Zhang, Y., Duchi, J. C. and Wainwright, M. J. (2013). Communication-efficient algorithms
for statistical optimization. {\it Journal of Machine Learning Research}, {\bf 14}, 3321-3363.

\item Zhao, T., Cheng, G. and Liu, H. (2016). A partially linear framework for massive
heterogeneous data. {\it Ann. Statist.}, {\bf 44}, 1400-1437.

\item Zou, H. and Yuan, M. (2008). Composite quantile regression and
the oracle model selection theory. {\it Ann. Statist.}, {\bf 36},
1108-1126.

\end{description}

\

\begin{figure}[htbp]
    \centering
\includegraphics{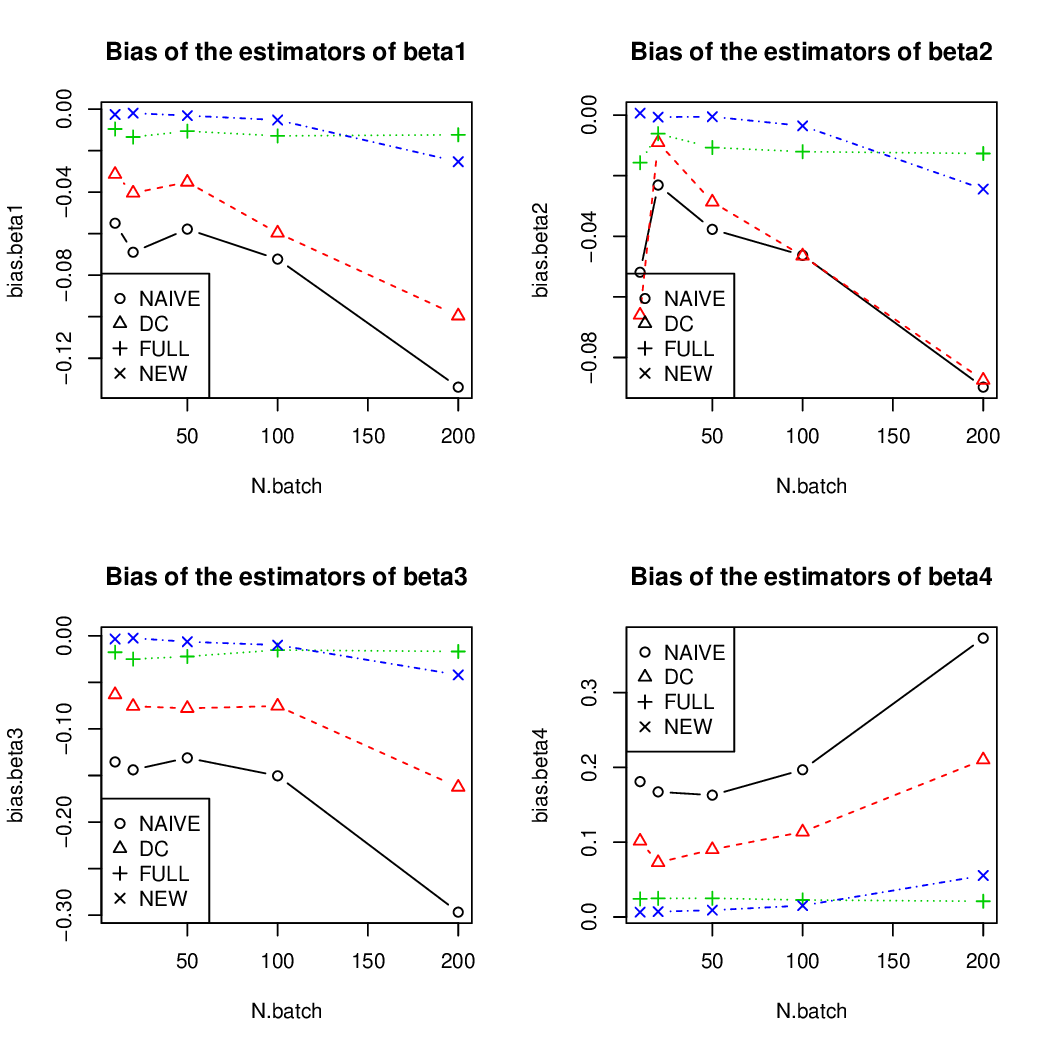}
      \caption{Bias of the estimators in Experiment 1}
      \label{fig1}

\end{figure}
\begin{figure}[htbp]
    \centering
  \includegraphics{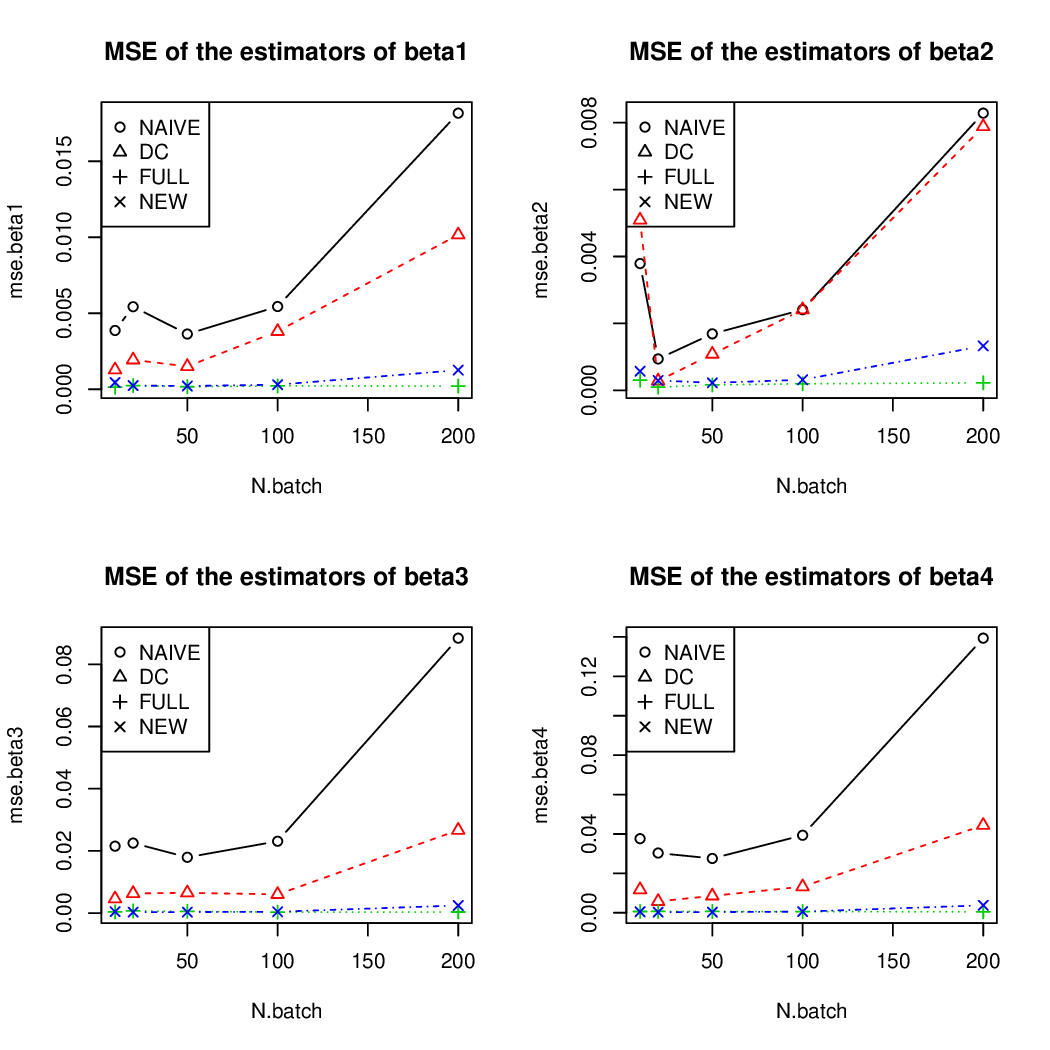}
      \caption{Mean square error of the estimators in Experiment 1}
      \label{fig1}

\end{figure}

\begin{figure}[htbp]
    \centering
\includegraphics{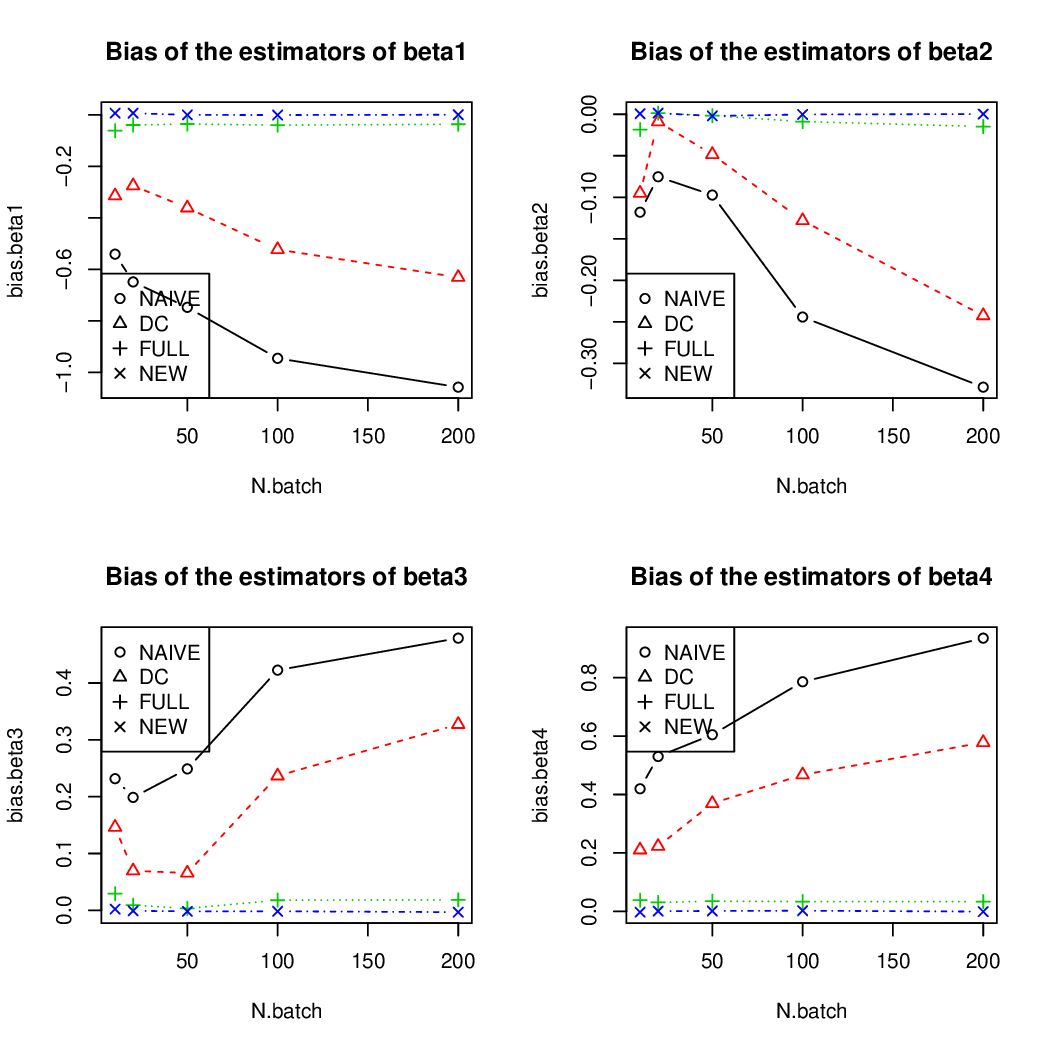}
      \caption{Bias of the estimators in Experiment 2}
      \label{fig1}

\end{figure}
\begin{figure}[htbp]
    \centering
\includegraphics{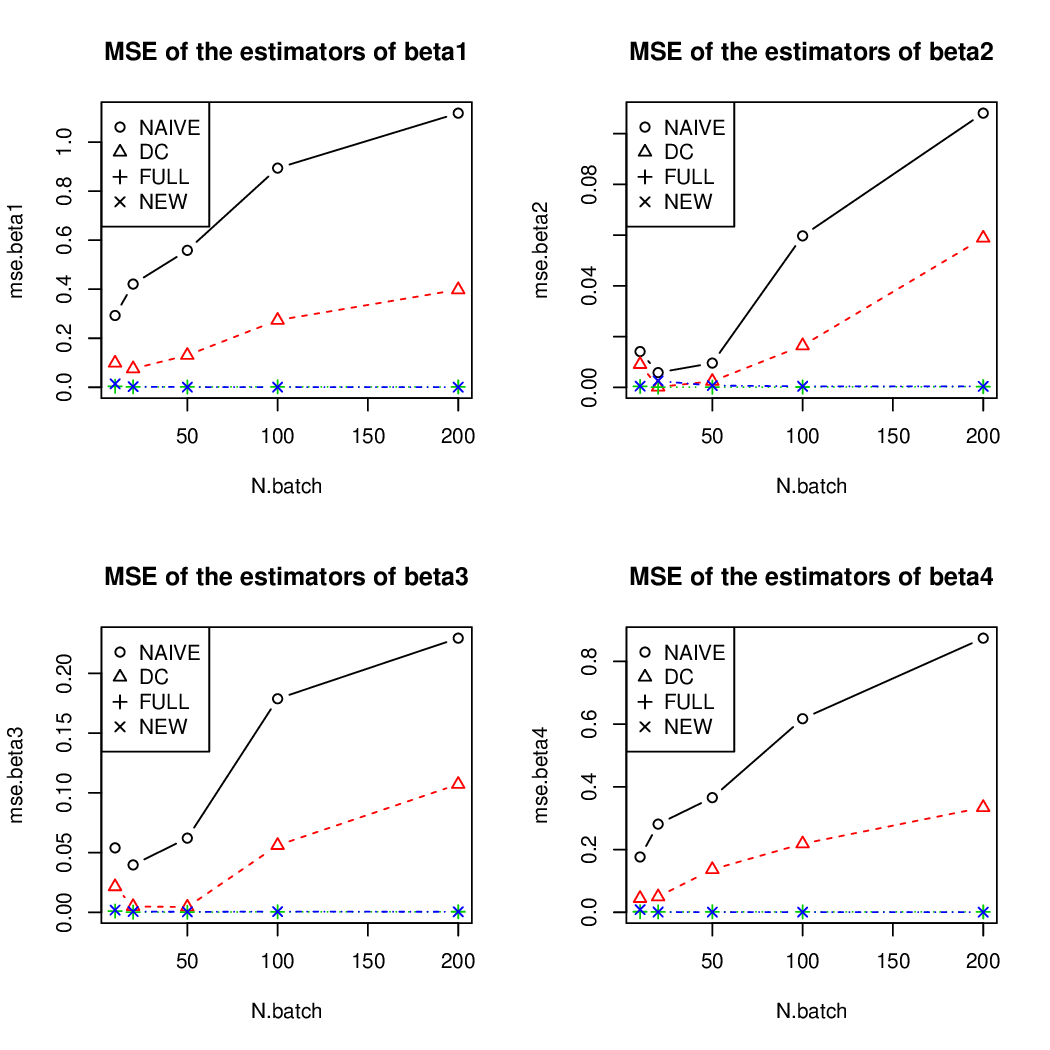}
      \caption{Mean square error of the estimators in Experiment 2}
      \label{fig1}

\end{figure}

\begin{figure}[htbp]
    \centering
\includegraphics{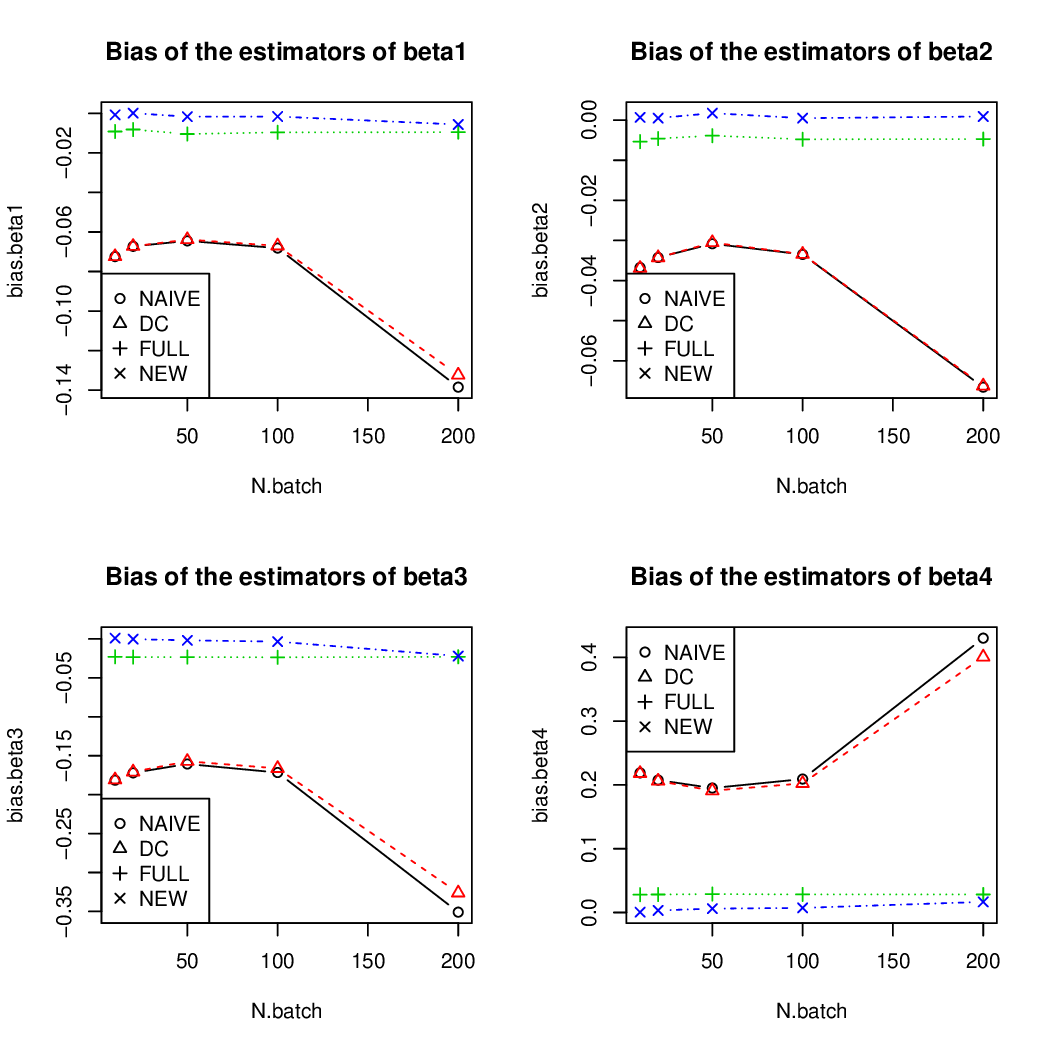}
      \caption{Bias of the estimators in Experiment 3}
      \label{fig1}

\end{figure}
\begin{figure}[htbp]
    \centering
  \includegraphics{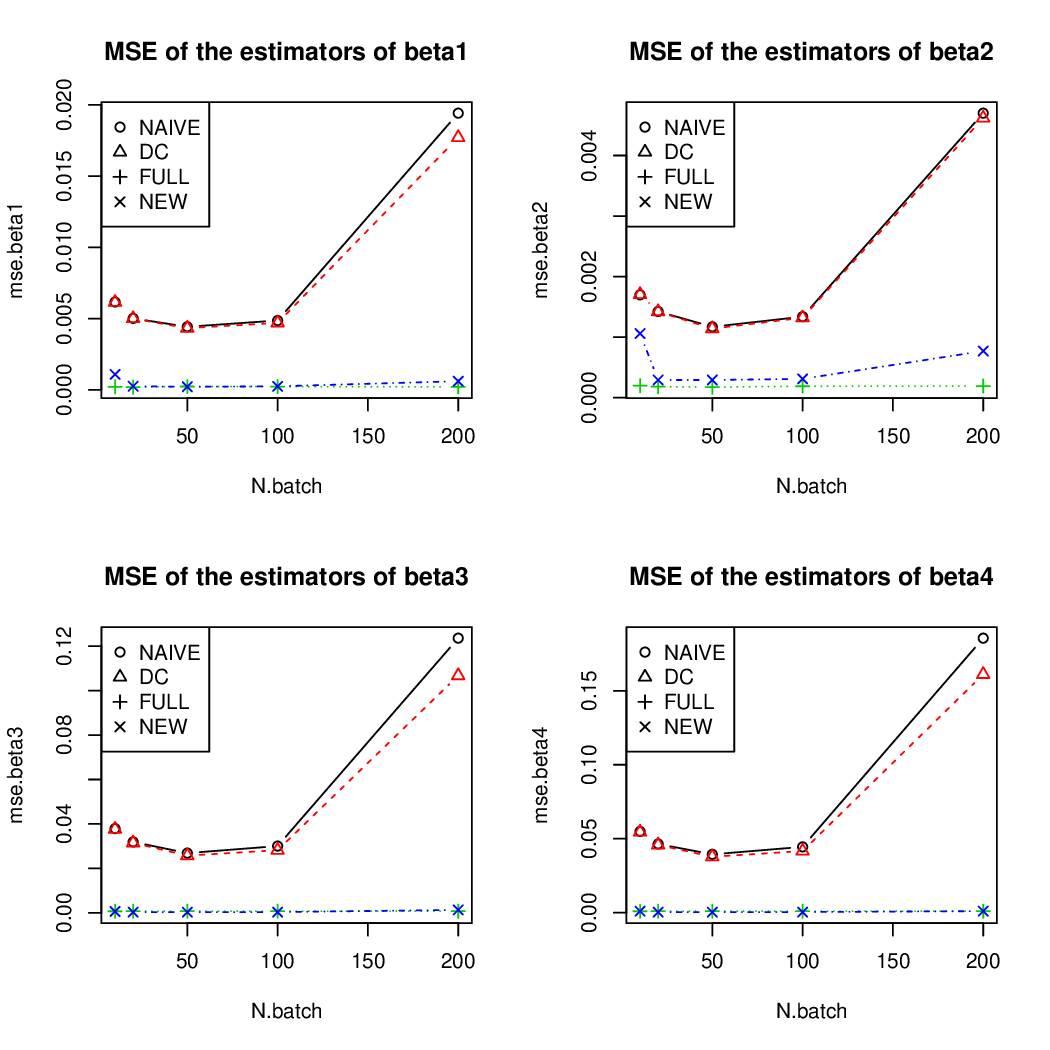}
      \caption{Mean square error of the estimators in Experiment 3}
      \label{fig1}

\end{figure}

\begin{figure}[htbp]
    \centering
\includegraphics{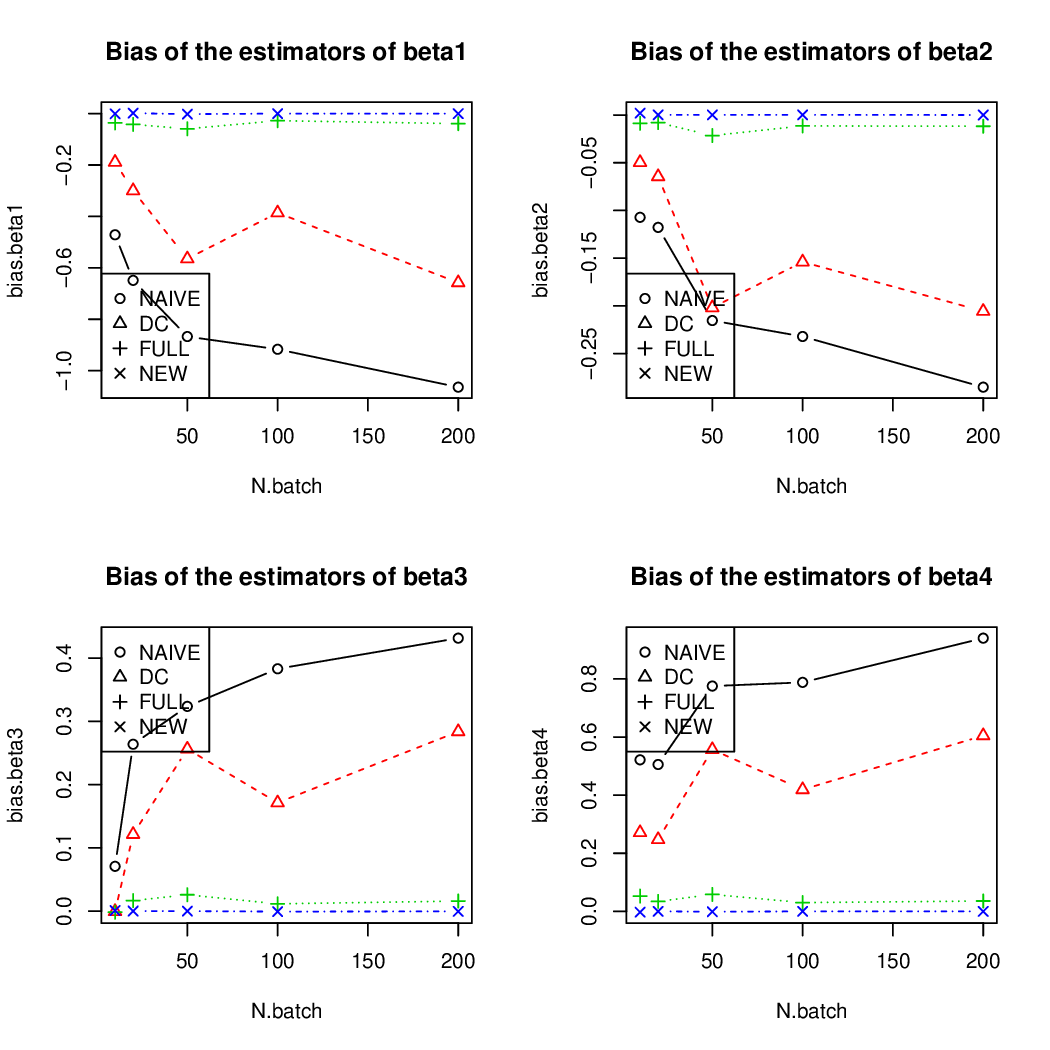}
      \caption{Bias of the estimators in Experiment 4}
      \label{fig1}

\end{figure}
\begin{figure}[htbp]
    \centering
  \includegraphics{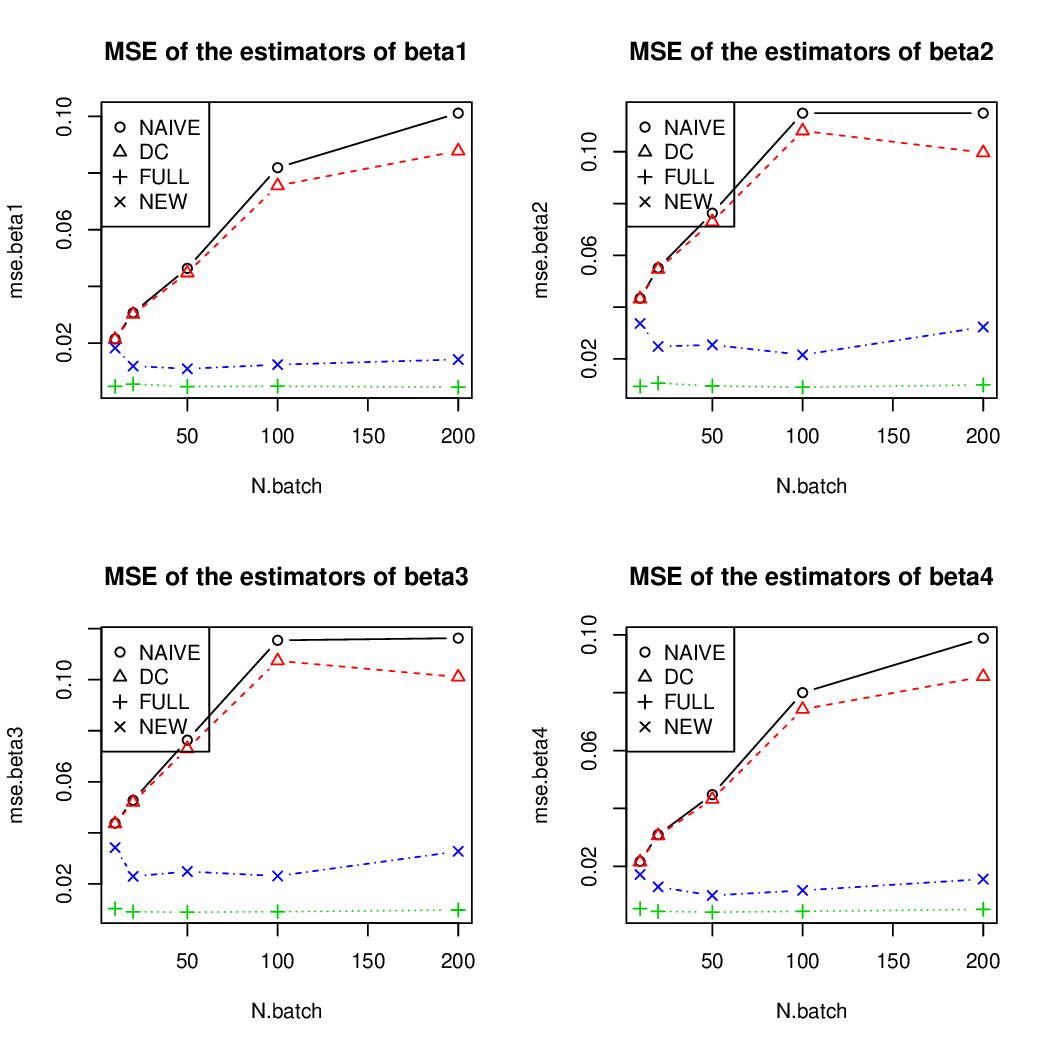}
      \caption{Mean square error of the estimators in Experiment 4}
      \label{fig1}

\end{figure}

\begin{figure}[htbp]
    \centering
  \includegraphics{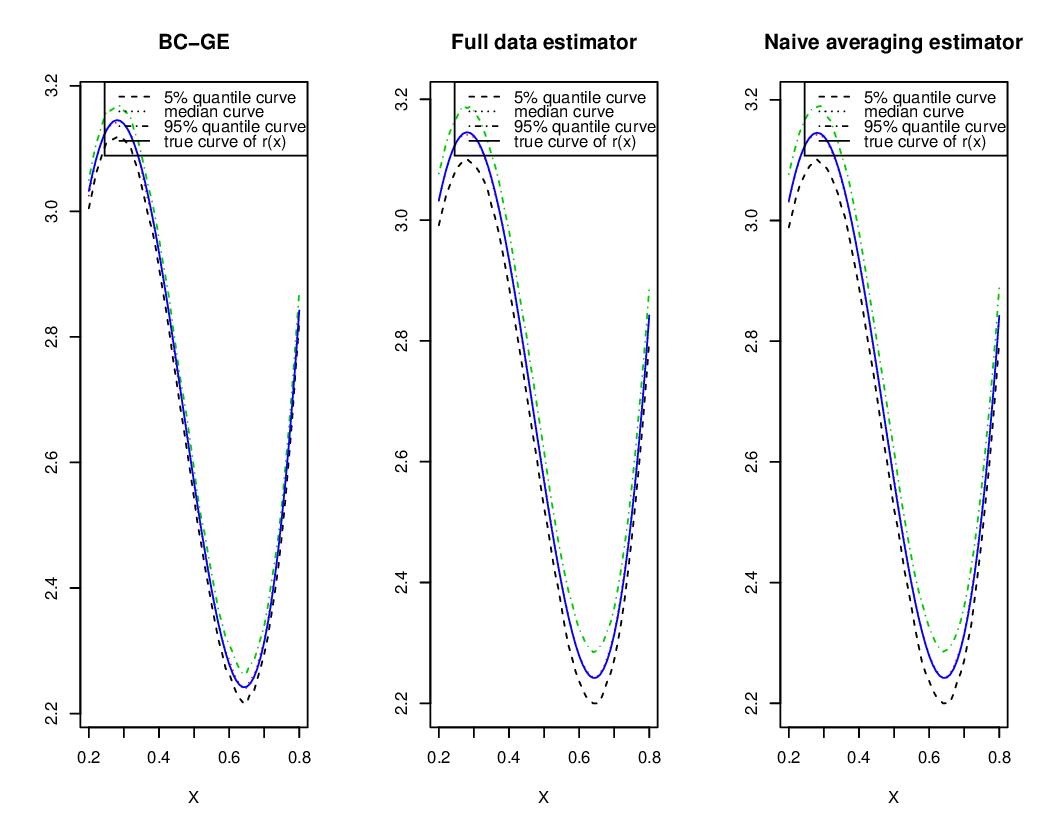}
      \caption{Quantile curves of the estimators in Experiment 5}
      \label{fig1}

\end{figure}

\newpage

\noindent
{\bf \large Appendix: Proofs}

\

\noindent {\it Proof of Lemma 3.1. } It is the direct result of the properties of expectation and variance of the original least squares estimation under linear model. $\square$

\

\noindent {\it Proof of Lemma 3.2. }
By the formula for the block matrix inversion, we have
\begin{eqnarray*}Var(\widetilde\theta^k|V_k)=\frac 1m\sigma^2\left(({\bf 1}^T{\bf 1})^{-1}+
({\bf 1}^T{\bf 1})^{-1}{\bf 1}^TV_k
D^{-1}V^T_k{\bf 1}({\bf 1}^T{\bf 1})^{-1}\right),\end{eqnarray*} where $D=V^T_kV_k-V^T_k{\bf 1}({\bf 1}^T{\bf 1})^{-1}{\bf 1}^TV_k$.
It follows from the definition of $V_k$ that
\begin{eqnarray*}&&
V^T_kV_k=\sum_{j=1}^N{\bf v}_{m}^k({\cal D}_j)({\bf v}_{m}^k({\cal D}_j))^T,
V^T_k{\bf 1}=\sum_{j=1}^N{\bf v}_{m}^k({\cal D}_j),\\&&
V^T_k{\bf 1}({\bf 1}^T{\bf 1})^{-1}{\bf 1}^TV_k=\frac{1}{N}\sum_{j=1}^N{\bf v}_{m}^k({\cal D}_j)\sum_{j=1}^N({\bf v}_{m}^k({\cal D}_j))^T,\\&&
V^T_kV_k-V^T_k{\bf 1}({\bf 1}^T{\bf 1})^{-1}{\bf 1}^TV_k=\sum_{j=1}^N{\bf v}_{m}^k({\cal D}_j)({\bf v}_{m}^k({\cal D}_j))^T-\frac{1}{N}\sum_{j=1}^N{\bf v}_{m}^k({\cal D}_j)\sum_{j=1}^N({\bf v}_{m}^k({\cal D}_j))^T.\end{eqnarray*}
The results above, Lemma 3.1
and Condition {\it C1} together imply that $({\bf 1}^T{\bf 1})^{-1}=N^{-1}$, $V^T_kV_k=O_p(\lambda^2N)$ and $V^T_k{\bf 1}=O_p(\lambda N)$. These result in $V^T_k{\bf 1}({\bf 1}^T{\bf 1})^{-1}{\bf 1}^TV_k=O_p(\lambda^2N)$, $D=O_p(\lambda^2N)$ and $V_k
D^{-1}V^T_k=O_p(N^{-1})$. Consequently, ${\bf 1}^TV_k
D^{-1}V^T_k{\bf 1}=O_p(N)$ and
$({\bf 1}^T{\bf 1})^{-1}{\bf 1}^TV_k
D^{-1}V^T_k{\bf 1}({\bf 1}^T{\bf 1})^{-1}=O_p(N^{-1}).$
Therefore, we have $Var(\widetilde\theta^k|V_k)=O_p(n^{-1})$.
The proof is completed. $\square$

\

\noindent {\it Proof of Theorem 3.3. } It is a direct result of Lemma 3.2. $\square$

\

\noindent {\it Proof of Theorem 3.4. } By the definition of the estimator, we have
$$(\widetilde\theta^k,\widetilde\xi^T)^T=(\theta^k,\xi^T)^T+\left(({\bf 1},V_k)^T({\bf 1},V_k)\right)^{-1}({\bf 1},V_k)^T\bm\epsilon^k,$$ where $\bm\epsilon^k=(\epsilon^k_1,\cdots,\epsilon^k_N)^T$. This shows that $(\widetilde\theta^k,\widetilde\xi^T)^T-(\theta^k,\xi^T)^T$ has mean zero and covariance $\frac 1m\sigma^2\left(({\bf 1},V_k)^T({\bf 1},V_k)\right)^{-1}$, and is normally distributed, asymptotically.
Thus, we only need to calculate the asymptotic variance of $\widetilde\theta^k$.

The proof of Lemma 3.2 and Condition {\it C3} indicate that
\begin{eqnarray*}\frac{1}{\lambda^2}D&=&\frac{1}{N\lambda^2}\sum_{j=1}^N{\bf v}_{m}^k({\cal D}_j)({\bf v}_{m}^k({\cal D}_j))^T-\frac{1}{N\lambda}\sum_{j=1}^N{\bf v}_{m}^k({\cal D}_j)\frac{1}{N\lambda}\sum_{j=1}^N({\bf v}_{m}^k({\cal D}_j))^T\\&\stackrel{p}\rightarrow& E[{\bf v}^k({\bf v}^k)^T]-E[{\bf v}^k]E[({\bf v}^k)^T.\end{eqnarray*} and moreover,
\begin{eqnarray*}Var(\sqrt n\,\widetilde\theta^k|V_k)&=&\sigma^2\left(1+
\frac{1}{N}\sum_{j=1}^N({\bf v}_{m}^k({\cal D}_j))^T
D^{-1}\frac{1}{N}\sum_{j=1}^N{\bf v}_{m}^k({\cal D}_j)\right)\\&\stackrel{p}\rightarrow& \sigma^2\left( 1+E[({\bf v}^k)^T]\left(E[{\bf v}^k({\bf v}^k)^T]-E[{\bf v}^k]E[({\bf v}^k)^T]\right)^{-1}E[{\bf v}^k]\right).\end{eqnarray*}
The proof is completed.
$\square$

\

{\it Proof of (\ref{(3.8)}).}
Similar (\ref{(2.2)}), the full data LASSO estimator of $\beta^k$ has the following representation:
$$\widehat\beta^k=\beta^k- ({\bf v}_n^k)^T\mbox{sgn}(\beta_S)+({\bf v}_n^k)^T\frac{1}{n}{\bf X}_{S}^T{\bm \varepsilon}.$$ Thus, its bias is
$-\lambda E[{\bf v}_n^k]\mbox{sgn}(\beta_S)$ and variance is
\begin{eqnarray*}Var(\widehat\beta^k)&=& (\mbox{sgn}(\beta_S))^TCov[{\bf v}_n^k]\mbox{sgn}(\beta_S)+\frac{1}{n}E\left[({\bf v}_n^k)^T\frac 1 n{\bf X}_{S}^T{\bf X}_{S}{\bf v}_n^k\right]\\&=&
 (\mbox{sgn}(\beta_S))^TCov[{\bf v}_n^k]\mbox{sgn}(\beta_S)+\frac{\sigma^2}{n}. \end{eqnarray*} Then, the mean square error of $\widehat\beta^k$ is
\begin{eqnarray*}&&MSE[\widehat\beta^k]\\&&= (\mbox{sgn}(\beta_S))^T E[{\bf v}_n^k](E[{\bf v}_n^k])^T\mbox{sgn}(\beta_S)+ (\mbox{sgn}(\beta_S))^TCov[{\bf v}_n^k]\mbox{sgn}(\beta_S)+\frac{1}{n}\\&&=(\mbox{sgn}(\beta_S))^T E[{\bf v}_n^k({\bf v}_n^k)^T]\mbox{sgn}(\beta_S)+\frac{\sigma^2}{n}.\end{eqnarray*}The proof is completed.
$\square$

\end{document}